# PBPU Elastomer Network Architecture Determination via Corresponding States Analysis of Mechanical Behavior

First Author: Sushanta Das

Department of Chemical Engineering, Indian Institute of Technology Bombay, Mumbai – 400 076, Maharashtra, India

Additional Director, Defence Research and Development Organization, New Delhi – 110011, India

ORCID: 0009-0006-6215-3573

Second Author: Hari Ramakrishna Sudhakar

Department of Chemical Engineering, Indian Institute of Technology Bombay, Mumbai – 400 076, Maharashtra, India

ORCID: 0009-0008-7215-1621

Corresponding author: Hemant Nanavati

Professor, Department of Chemical Engineering, Indian Institute of Technology Bombay, Mumbai – 400 076, Maharashtra, India

ORCID: 0000-0002-5982-6531

*E-mail: hnanavati@iitb.ac.in, hnanavati@che.iitb.ac.in

## Abstract

In this work we examine the effect of $R$=[NCO]/[OH] in the $R\leq 1$ regime, on the resultant structural topology of polybutadiene polyurethane (PBPU) elastomer networks based on hydroxy-terminated polybutadiene (HTPB). We employ stress-elongation ($\sigma - \lambda$) behavior and its modeling, as a tool. We examine this property via a combination of our model for the finite chain phantom networks incorporating the HTPB structural information, with the slip-tube model from the literature, suitably modified phenomenologically. We implement a further normalized Mooney-Rivlin (MR) representation (corresponding deformation states plots), to remove any magnitude bias on the model parameters. The now revealed curvatures of all the MR plots, in turn, reveals the non-correlation between the chain size and crosslink density. This discrepancy occurs due to the $R$-dependent majority presence of network defects due to sol effects (as obtained from swelling experiments) and non-load bearing pendant branches on the load-bearing network chains.

# 1 Introduction

In this work, we examine the stoichiometry-dependent stress-elongation relationships for PBPU elastomer networks based on hydroxy terminated polybutadiene (HTPB). While cure-dependent elasticity relationships have been examined previously (e.g., the work of Klueppel *et al.*, [1–6]), those networks were synthesized for a controlled variation of the crosslinking agent such as Sulphur or dicumyl peroxide. In our work, this relationship is obtained indirectly, by controlling the stoichiometry, i.e., the ratio, R=[NCO]/[OH]. Elasticity analyses reported previously (e.g., the works of Sekkar *et al.*, [7–10]) consider $R>1$. The crosslink density is estimated via various methods such as the Mooney-Rivlin (MR) parameters with auxialliary data from equilibrium swelling experiments.

We consider $R\leq 1$, and map the stoichiometry to the structural topology via stress-elongation behavior. The stress-elongation ($\sigma-\lambda$) experiment data are examined in the more informative MR form. Across $R$, these data span an order of magnitude, which conceals the MR features of the softer materials. We mathematically expose these features by a corresponding strain states method; i.e., normalizing the data by a reference within the corresponding MR plot itself.

We recognize the experiment protocol causes deliberate errors, particularly at low $\sigma-\lambda$. While these may be disregarded for stiffer materials, and for performance evaluation, they confound structural analyses of our materials. Therefore, our analysis includes a systematic and rational mitigation of these errors.

We model our refined MR data, with a combination of (i) mathematically accurate framework for ideal freely jointed chains (FJC) for the chemically crosslinked component, (ii) deformation dependent ideal junction fluctuations, and (iii) carefully selected literature model for the entanglement component, that is suitably modified to represent the observed phenomena. We incorporate the structural information of our PBPU elastomers and infer stoichiometry-dependent topology of the resultant structures. Swelling experiments provide the sol fraction, which does not bear load. By accounting for the presence of defects in a rational manner, this analysis relates $R$ to the molecular morphology of PBPU elastomers.

# 2 Materials

The raw materials employed for the sample preparation are listed in 0.

The properties of the HTPB (obtained as previously reported) [7–10] are listed in Table 2.

**Table 1.** Structures of various raw materials used for the preparation of PBPU elastomers, along with their schematic representations.

| **Reactants** | | **Schematic** |
|---|---|---|
| 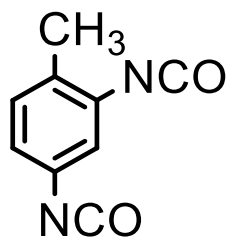  2,4-Toluenediisocyanate (2,4-TDI) (80%) | 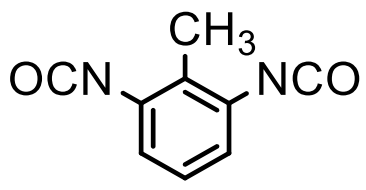  2,6-Toluenediisocyanate (2,6-TDI) (20%) | 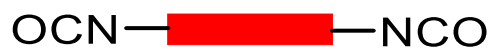  |
| 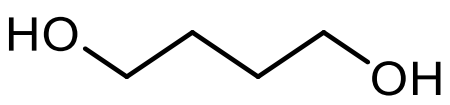  n-Butanediol (nBD) | | 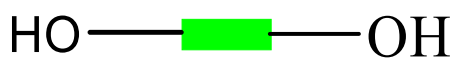  |
| 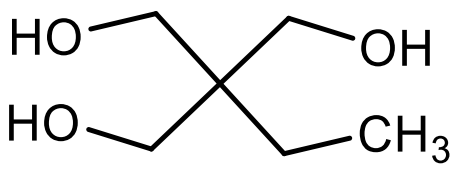  Trimethylolpropane (TMP) | | 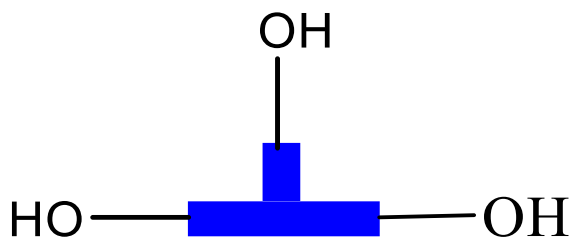  |
| 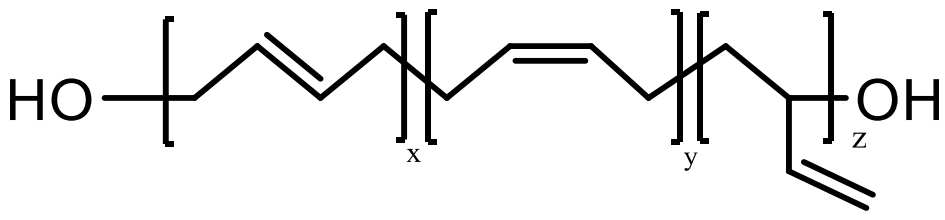  Hydroxyl Terminated Polybutadiene (HTPB) | | 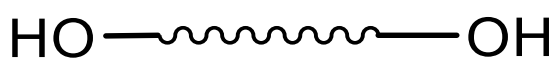  |

**Table 2.** Properties of HTPB as determined by various chemical analysis and instrumental techniques (as indicated in the table).

| **Property** | **Value** |
|---|---|
| Hydroxyl value (mg KOH/g HTPB) | 41.30 |
| Molecular Weight (Mn by VPO) | 2850 |
| Polydispersity Index (PDI by GPC) | 1.85 |
| Moisture Content (%) | 0.05 |
| Viscosity at 30°C (cP) | 5909 |
| Volatile Matter (%) | 0.18 |
| Average functionality ($f_n$) | 2.1 |
| Glass Transition Temperature ($T_g$) (°C) | -78 |

We examine five $R$=[NCO]/[OH] stoichiometries (A to E, respectively), ranging from 1.0 to 0.8. The aborted sixth stoichiometry, $R$=0.75, does not yield a percolated network, and is not useable for our analysis.

The compositions in our formulations are nBD and TMP in the ratio, ~2:1, in terms of the equivalent OH, for all $R$ values; i.e., there are ~2 OH from nBD for every OH from TMP; thus, the molar ratio of nBD and TMP is 3:1. Considering $R$=1 as the benchmark, there are 100/9 NCO equivalents (number of NCO from TDI) for each OH equivalent from TMP. Then, the molar ratio of TDI and TMP is 50/3:1. The remaining OH are sourced from HTPB to achieve perfect stoichiometry. While formulating the various stoichiometries, the combined weight of HTPB and TDI is held constant. Then, considering the $R$=1 stoichiometry described above as the basis (1/3 mmole of TMP), there are 5.176 $R$ + 0.3749 mmoles of TDI are in the formulation, with 2.5237 $R^2$ – 4.6514 $R$ + 13.177 g of HTPB. We prepare samples PBPU-A ($R$=1), PBPU-B ($R$=0.95), PBPU-C ($R$=0.90), PBPU-D ($R$=0.85), and PBPU-E ($R$=0.80).

The microstructure details of the HTPB are obtained via $^1$H-NMR (operating frequency: 500 MHz, internal reference: tetramethylsilane, solvent: $CDCl_3$). They are listed in Table 3.

**Table 3.** Details of functionality and backbone microstructure of the free radically produced HTPB from $^1$H-NMR spectroscopy.

| | | | Content (%) |
|---|---|---|---|
| 1. | Functionality microstructure | Cis/G | ~13 |
| | | Trans/H | ~60 |
| | | Vinyl/V | ~27 |
| 2. | Backbone microstructure | Cis | ~20 |
| | | Trans | ~60 |
| | | Vinyl | ~20 |

The reaction between the OH species and the TDI was carried out at 50°C in a glass reactor setup. The product was cast on Teflon-coated SS trays. Further curing (crosslinking) was carried out over 1 day at room temperature and 5 days at 50°C. The sheets were conditioned for 2 weeks in terms of the RH, and cut into dumbbells as per ASTM D412.

## 3 Experimental Methods

In this work, we examine PBPU elastomer networks via stress-elongation experiments and their analyses. Swelling experiments provide information on the sol fraction, which does not bear load, and needs to be accounted for.

Uniaxial testing of the dumbbell samples was carried out in Tinius Olsen UTM at crosshead speed of 50 mm/min with a sample gauge length of 25 mm, at 24°C. We employed a 100 N load cell with an accuracy of 0.01 N. After the set preload of 0.1 N, the experimental procedure involves simultaneous recording of the crosshead position, $X$, the load in N, the strain (via a laser extensometer), ε, based on the gauge length being 25 mm. The conventional procedure is to subtract this preload, and shift the axis. The laser extensometer requires that the very narrow reflecting adhesive silver strips be stuck and wrapped around the sample, separated by the gauge length. A typical stress-elongation ($\sigma - \lambda$) plot is provided in Figure 1(a). The description of the $\sigma - \lambda$ relationship needs to be normalized, to remove the bias towards the high $\sigma$ at high $\lambda$. This is achieved by analyzing the Mooney-Rivlin (MR) plot, $\sigma_{red}$ vs $\frac{1}{\lambda}$; the reduced stress, $\sigma_{red} = \frac{\sigma}{\left(\lambda - \frac{1}{\lambda^2}\right)}$ (Figure 1(b)) [11–17].

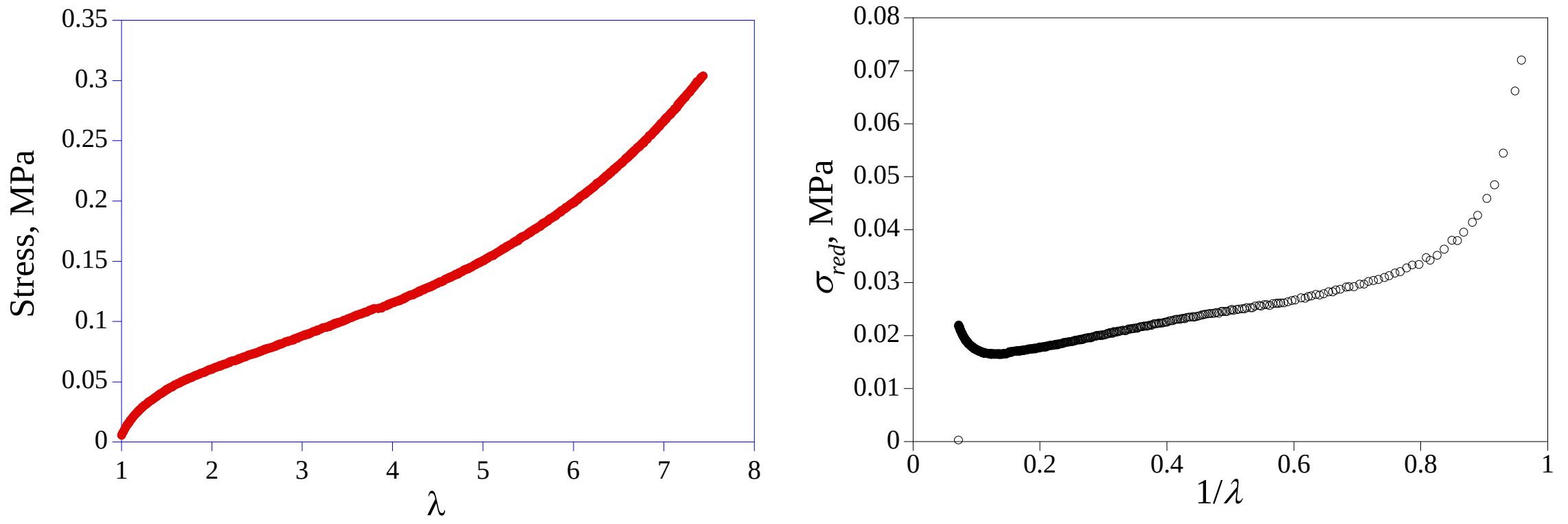

**Figure 1.** Elasticity plots for PBPU elastomers, for R=0.8 (a) Typical $\sigma$-$\lambda$ plot (b) MR plot of $\sigma_{red}$ vs $\frac{1}{\lambda}$

There are a few concerns with this established protocol. These are listed and addressed in Appendix A. The resultant stress-elongation plot, employing the original and corrected $\lambda$ and the magnified plot in the low $\lambda$ region, are depicted in Figure 2. This enables an accurate development of the MR plot (Figure 3).

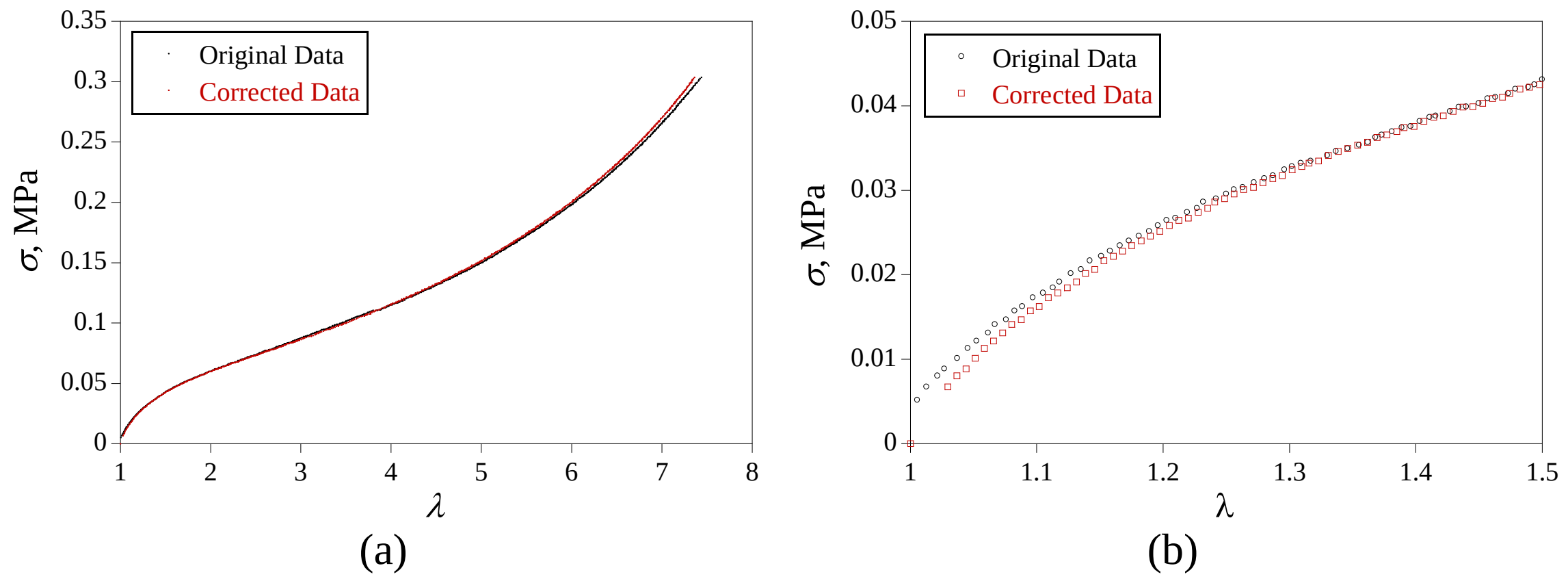

**Figure 2.** Comparison of the original $\sigma-\lambda$ data and those modified by the $l_e/X$ analysis. (a) data over the entire range exhibits no visible change because of the correction (b) data over a 50% strain range exhibits a very slight variation due to the correction up to 20% strain, and negligible difference beyond 20% strain.

Hence, we find that even though the $\sigma - \lambda$ plots do not exhibit visible variation on correction, the MR plots indicate a shift towards lower elongation and higher initial $\sigma_{red}$. Another important observation is that the MR plot exhibits an upturn at low tensile $\lambda$, suggesting that there might be an exponential variation in the entanglement modulus. The corrected upturn at lower $\lambda$, however, results in lower $\sigma_{red}$.

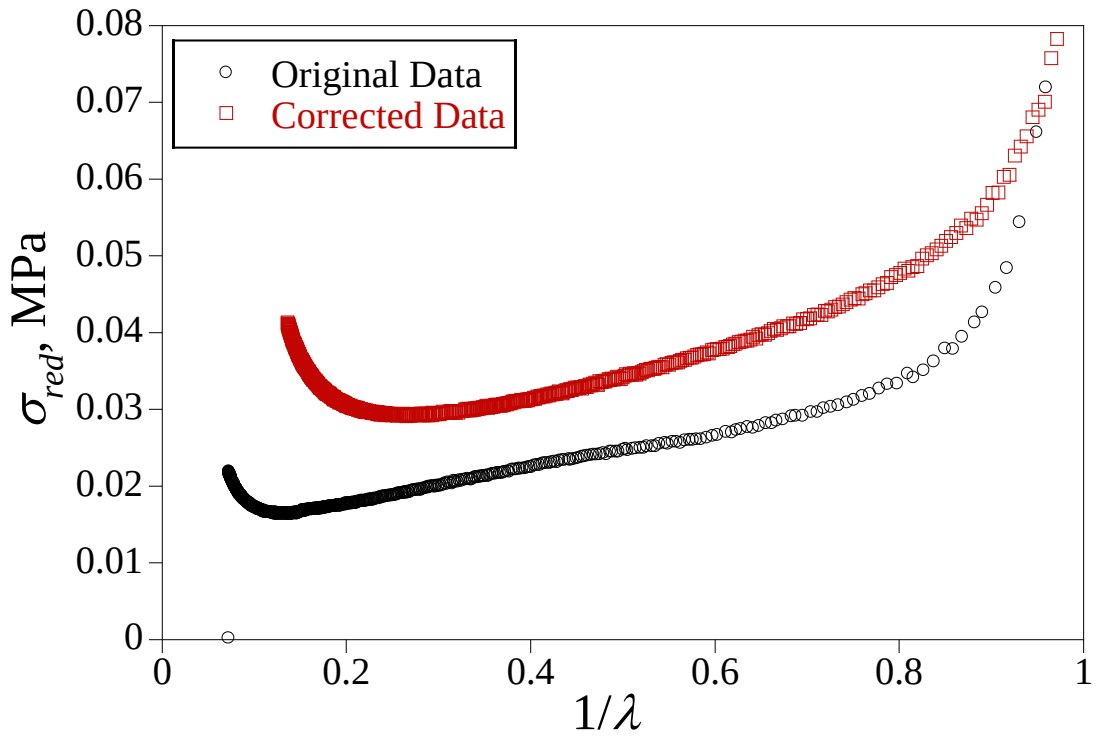

**Figure 3.** The effect of the correction of the $\sigma - \lambda$ data via the $l_e/X$ analysis indicates significant variation on the MR plot.

The MR data for all stoichiometries are shown in Figure 4(a). $\sigma_{red}$ traverses an order of magnitude as R increases from 0.8 to 1.0. In order to depict the relative variation in $\sigma_{red}$ with $\frac{1}{\lambda}$, Figure 4(b) provides $\sigma_{red}$ on a Log scale. We find that qualitatively, the relative decrease in $\sigma_{red}$, is similar across all stoichiometries, and minima are visible in chemistries C – E, with a finite extensibility upturn at low $\frac{1}{\lambda}$ (shown for R=0.8 network in Figure 3).

Consistent with our observations, literature experiments also indicate an upturn in the MR plot (concavity upwards), for $0.7<\frac{1}{\lambda}<0.95$ [1–4,12,18,19] (data for $0.95<\frac{1}{\lambda}<1.0$ are not considered reliable due to the errors in estimating $\lambda$ in this regime). This upturn in the tensile regime, $\frac{1}{\lambda}<1.0$, is clearly visible even in experiments spanning both regimes – tension and compression [19–21]. The cause of this upturn is not clear, although some researchers have attributed it to temporary entanglements. Others [12,18] have attributed it to artefacts, and have disregarded the upturn.

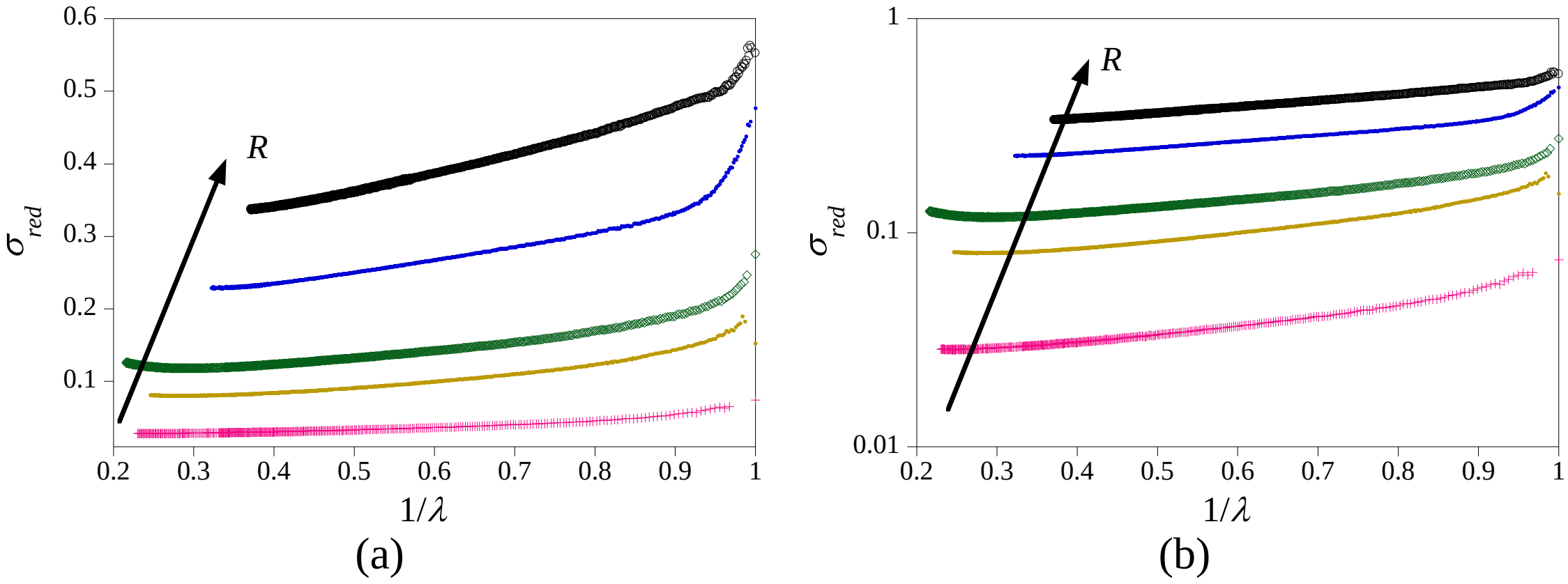

**Figure 4.** MR plots with representative samples of all stoichiometries, A–E, with the direction for increasing *R* indicating the respective plot for each stoichiometry. (a) $\sigma_{red}$ traverses an order of magnitude as *R* increases from 0.8 to 1.0. (b) Representing $\sigma_{red}$ on a log scale indicates that the relative variation with deformation, is qualitatively similar for all *R* values. The variation of $\sigma_{red}$ across the range of *R* approximately one order of magnitude.

The equilibrium swelling experiment is a very common tool for estimating crosslink density. The experimental method [7–10,22–24] is elaborated in Appendix B. We have considered generalized functionality dependent effects [25–29]; in our networks, functionality, *f*=3. The framework considers both the affine network model and the phantom network model. The fundamental finding of our swelling analyses is that there is a sol fraction, $f_{\text{sol}}$, which varies from ~0.11 at *R*=1 to ~0.41 at *R*=0.8. We describe next setting up of the stress-elongation modelling framework which can be implemented self-consistently with the swelling studies.

# 4 Stress-Elongation Data Modeling Development

We divide the modeling framework in two steps. In the first step, we describe the modeling of the phantom network of finite chains, for PBPU elastomers. In the second step, we include the effects of entanglements and then model-fit the combined MR behavior of each *R* network. Subsequently, the swelling experiments results will be incorporated self-consistently.

## 4.1. Conventional Stress-Elongation Modeling

### Chemical Crosslink Stress Component

We begin with the framework for a phantom network of finite chains [30]. The strain energy density of a network is $N\,\mathrm{k}TW_{def}$. For an ideal network (defect-free, monodisperse, with no dangling ends) $N\,\mathrm{k}T = \dfrac{\rho \mathrm{R} T}{M_X}$. $M_X$ is the molecular weight of the chain between crosslinks, $\rho$ is the density, R is the universal gas constant, $T$ is the absolute temperature. For a chain contains $n$ segments, each of mass $M_S$, $M_X = nM_S$. In addition to $N\,\mathrm{k}T$, the expression for $W_{def}$, which is a function of $\Lambda = \dfrac{1}{3n}\left(\lambda^2 + \dfrac{2}{\lambda}\right)$, also contains $n$. The stress-elongation relationship is obtained from:

$$\sigma = \frac{\rho \mathrm{R} T}{nM_S}\frac{dW_{def}}{d\lambda} = \frac{\rho \mathrm{R} T}{nM_S}\frac{dW_{def}}{d\Lambda}\frac{d\Lambda}{d\lambda} = \frac{\rho \mathrm{R} T}{nM_S}\frac{2}{3n}\left(\lambda - \frac{1}{\lambda^2}\right)\frac{dW_{def}}{d\Lambda} \quad (1)$$

$$W_{def} = \left(\frac{n + 0.5n^{0.75}}{2}\right)\left(\Lambda - \frac{1}{n}\right) + \left(\frac{3 + 0.5n^{0.75} - 2n}{2}\right)\ln\left(\frac{1-\Lambda}{1-\frac{1}{n}}\right) \quad (2)$$

Then

$$\sigma = \frac{\rho \mathrm{R} T}{nM_S}\frac{2}{3n}\left(\lambda - \frac{1}{\lambda^2}\right)\left(\frac{n + 0.5n^{0.75}}{2} + \frac{\left(3 + 0.5n^{0.75} - 2n\right)}{2\left(1-\Lambda\right)}\right) \quad (3)$$

Next, we account for the effects of junction fluctuations. The fluctuations can be modeled as ideal FJC's of $n_f$ segments, where $n_f$ is a function of deformation, $\lambda$; i.e.,

$$n_{eff} = n\frac{f - \Lambda}{f - 2 + \Lambda} \Rightarrow \frac{1}{n_{eff}} = \left(1 - \frac{2(1-\Lambda)}{f - \Lambda}\right)\frac{1}{n} \quad (4)$$

However, the breadth of the fluctuation is greater than that of an ideal FJC. Hence, we can also approximate the fluctuation to be Gaussian, instead of an ideal FJC. For such Gaussian chains as well, $n_f$ would remain the same function of $\lambda$. We incorporate eqn. (4) into $N\,\mathrm{k}T$, i.e, the crosslink modulus becomes

$$G_X = \frac{\rho \mathrm{R} T}{nM_S}\left(1 - \left(\frac{2(1-\Lambda)}{f - \Lambda}\right)\right)\Bigg|_{\Lambda = \frac{1}{n}} = \left(\frac{\rho \mathrm{R} T}{nM_S}\right)\left(\frac{n+1}{3n-1}\right) \quad (5)$$

i.e., the phantom network stress,

$$\sigma_{Ph} = \frac{\rho \mathrm{R} T}{n M_S}\left(1 - \left(\frac{2(1-\Lambda)}{f-\Lambda}\right)\right)\frac{2}{3n}\left(\lambda - \frac{1}{\lambda^2}\right)\left(\frac{n + 0.5n^{0.75}}{2} + \frac{\left(3 + 0.5n^{0.75} - 2n\right)}{2(1-\Lambda)}\right) \quad (6)$$

Then, since the strain energy density is reduced by fluctuations,

$$W_{Ph,def} = W_{def} - W_f$$

where

$$W_f = \frac{2}{n}\left(\frac{1-\Lambda}{f-\Lambda}\right)\frac{3n}{2}\Lambda \quad (7)$$

We can consider $G_s \sim nG_X$ to be “segment” based shear modulus. In the literature, $G_X$ and $n$ are lumped because those frameworks consider only Gaussian networks [12,13,18,31], or they are combined self-consistently with approximations of the infinite Taylor’s series of the inverse Langevin function [1–6,11,32–37].

The effect of the fluctuations is to decrease the $N\mathrm{k}T$ strain energy density pre-factor for Gaussian networks. For our finite chain phantom networks,

$$\frac{G_s}{n}W_{Ph} = \frac{G_s}{n}\left(\left(\frac{n + 0.5n^{0.75}}{2}\right)\left(\Lambda - \frac{1}{n}\right) + \left(\frac{3 + 0.5n^{0.75} - 2n}{2}\right)\ln\left(\frac{1-\Lambda}{1-\frac{1}{n}}\right) - \left(\frac{1-\Lambda}{f-\Lambda}\right)3n\Lambda\right) \quad (8)$$

We consider $f = 3$ because all crosslinks are trifunctional – whether they arise from TMP or from pendant OH on HTPB. Then, based on the last term of equation (8) and its derivative, the stress is reduced due to junction fluctuation by

$$\sigma_f = \frac{\rho \mathrm{R} T}{n M_S}\frac{2}{3n}\left(\lambda - \frac{1}{\lambda^2}\right)\left(\frac{3\left(\Lambda^2 - 6\Lambda + 3\right)}{(\Lambda - 3)^2}\right). \quad (9)$$

Thus for finite chain phantom networks, the stress becomes

$$\sigma_{Ph} = \frac{\rho \mathrm{R} T}{n M_S}\frac{2}{3n}\left(\lambda - \frac{1}{\lambda^2}\right)\left(\left(\frac{n + 0.5n^{0.75}}{2} + \frac{\left(3 + 0.5n^{0.75} - 2n\right)}{2(1-\Lambda)}\right) - \left(\frac{3\left(\Lambda^2 - 6\Lambda + 3\right)}{(\Lambda - 3)^2}\right)\right) \quad (10)$$

Then

$$\sigma_{Ph,red} = \frac{G_s}{n}\frac{2}{3n}\left(\left(\frac{n+0.5n^{0.75}}{2}+\frac{\left(3+0.5n^{0.75}-2n\right)}{2\left(1-\Lambda\right)}\right)-\left(\frac{3\left(\Lambda^2-6\Lambda+3\right)}{\left(\Lambda-3\right)^2}\right)\right) \tag{11}$$

In order to estimate $G_s$, it is necessary to identify the segment, and then compute the mass per segment. The Kuhn length of a segment of HTPB will need to represent the three components, *cis*, *trans* and *vinyl*. The characteristic ratios of the *cis* and *trans* are 4.9 and 5.8 respectively [38,39]. The characteristic ratio for the vinyl component would be similar to that for poly(1-butene), i.e., ~11 [40]; Zhou *et al.* [41], have reported values ~10 for atactic poly(1,2-butadiene). The fully stretched distances for poly(cis-butadiene) and poly(trans-butadiene) monomers are 4.56Å and 5.08Å, respectively [42]. The fully stretched length for the vinyl monomer is ~2.5Å. Considering the definition of the Kuhn segment in terms of its $\langle h_0^2 \rangle = na^2$ and its $h_{\max} = na$ (as well as the relative fractions of the corresponding components from Table 3), there are ~2.5 monomers per Kuhn segment. Thus, the Kuhn segment length, *a*~11.1 Å, $M_s \sim 135$. Since the densities of the samples are ~0.93g/cc, $G_s \sim 17\,\text{MPa}$.

The fluctuations of the phantom network junctions which lead to stress reduction by $\sigma_f$, are suppressed by the entanglements of the surrounding chains. Therefore, on this finite chain phantom network model, we superimpose the effect entanglements.

**Entanglement Stress Model Selection**

The entanglement models reported include Edwards' slip-link model [43,44], Heinrich's tube model [37,45], with Klüppel's extension [1–6,36], the slip-tube model [12,18,31,46], the non-affine network model [33], the network averaging model [35], and a general constitutive model [34]. These are consistent with the literature data that the combined MR plot for $\sigma_{red}$ in tension and compression, exhibits a maximum in the compression region. We select here, the slip-tube model (eqn. (12)), which has been explicitly compared with the combined experimental MR data across tensions and compression; it exhibits a maximum at $\frac{1}{\lambda}$~1.8 [12,19,21,31]. Unlike the frameworks of Edwards, Heinrich and Klüppel, which self-consistently combine the entanglement effects with the crosslink effects, the framework explicitly adds the entanglement stress, to the phantom network stress. We note that these frameworks also exhibit a maximum in the compression region, with the location of the maximum dependent on the model parameters.

$$\sigma_e = \frac{G_e\left(\lambda - \frac{1}{\lambda^2}\right)}{-0.35 + 0.61(\lambda)^{-0.5} + 0.74\lambda} \tag{12}$$

Then the total stress, $\sigma = \sigma_{Ph} + \sigma_e$. For the MR analysis,

$$\sigma_{e,red} = \frac{G_e}{-0.35 + 0.61(\lambda)^{-0.5} + 0.74\lambda} \tag{13}$$

When combined with eqn. (11),

$$\sigma_{red} = \sigma_{Ph,red} + \sigma_{e,red} \tag{14}$$

### 4.2. Framework Development for PBPU Network Analyses

**Corresponding States via Self-scaling of MR Plots**

Equation (13) is consistently convex upwards, particularly for Gaussian networks. However, for finite networks (*n*~60), $\sigma_{red}$ exhibits an inflection at $\frac{1}{\lambda}$~0.56, if $G_c \sim G_e$. Also, our data exhibit concavity upwards, in their MR plots in the range, $\frac{1}{\lambda}$>0.7.

In order to better visualize the data simultaneously across *R*, we consider $\frac{1}{\lambda} = 0.7$ as the reference state and convert MR plots of all compositions to a common "corresponding states" scale; i.e., for each data set, we normalize $\sigma_{red}$ by its *own* reference state, $\sigma_{red}\big|_{\frac{1}{\lambda}=0.7}$. Thus we develop a corresponding states scale for $\sigma_{red}$. Then, $\sigma_{red}^{norm}\big|_{\frac{1}{\lambda}=0.7} = 1$, for all compositions. Such self-scaling exposes the otherwise suppressed (Figure 4) curvatures of the data across all *R* values. Figure 5 depicts representative corresponding states data (LOESS-smoothened for clarity) of the five stoichiometries.

The curves exhibit significant similarities in their curvature. *R*≤0.9 plots exhibit clear minima. For *R*≥0.95, the samples fail before the MR plots reach their mathematical minima, although there is an onset towards the upturn. In the range, $\frac{1}{\lambda}$>0.7, the self-scaled plots for *R*≥0.9 are nearly superimposed until the upturn onset. The lowest values of $\sigma_{red}^{norm}$ as well as their $\frac{1}{\lambda}$ locations, can be traced to increase with *R*. The locations increase with decreasing *R* from $\lambda \sim 2.5$ to $\lambda \sim 3.75$, i.e., increasing by a factor of 1.5. Qualitatively, the ratio, $\frac{n(R\sim 0.8)}{n(R\sim 1.0)} \sim \left(\frac{3.75}{2.5}\right)^2 \sim 2.25-3.0$. This is in contrast to the corresponding lowest $\sigma_{red}$ values varying from 0.35 to 0.028, decreasing by a factor of ~12.5. For a conventional network where the high MW chains are crosslinked, $\sigma_{red} \propto \frac{\rho \mathrm{R} T}{n M_S}$; then $\frac{n(R\sim 0.8)}{n(R\sim 1.0)} \sim \left(\frac{0.35}{0.028}\right) \sim 12.5$. Thus, the self-scaling step, suggests non-applicability of conventional elastomer modeling.

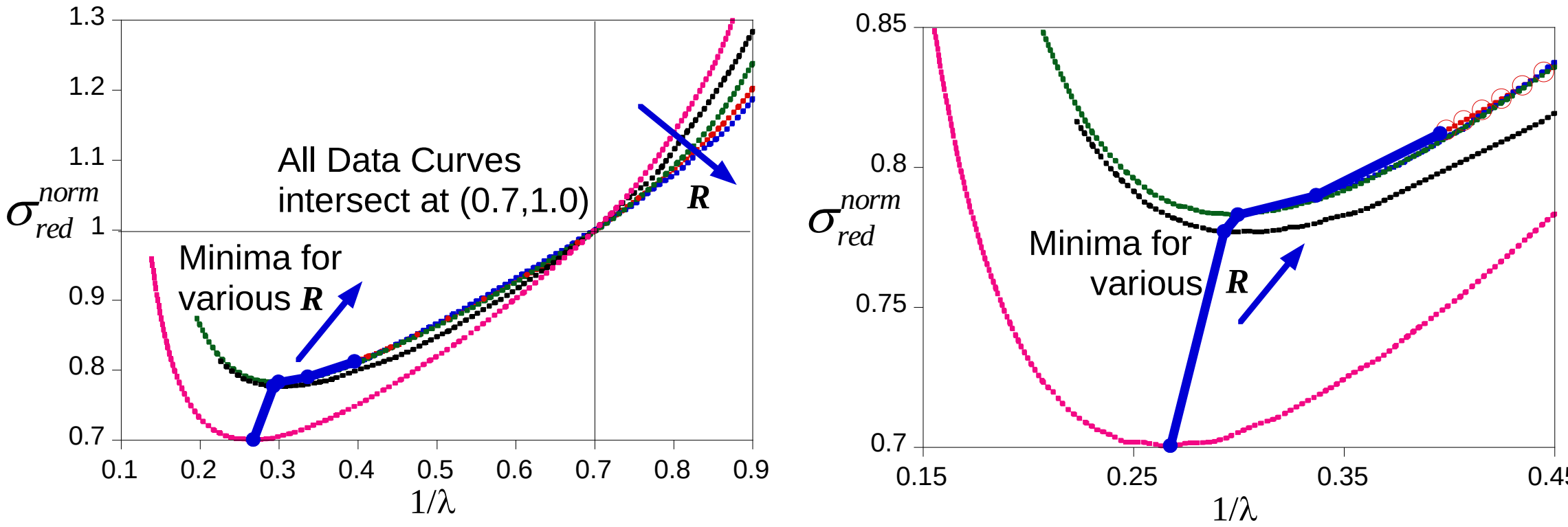

(a) (b)

**Figure 5.** Corresponding States MR plots for PBPU networks from the 5 stoichiometries. The heavy blue line with markers indicates the locations of the minima for the five $R$ values. For $R$=1.0 and $R$=0.95, the samples fail before the plots reach their MR minima. The variation in the location of the minima is indicated by a heavy dark line with visible makers (a) The direction of increasing $R$ reverses across the intersection at the reference point (b) The plots zoomed in the neighborhood of the minima indicate that the minima $\frac{1}{\lambda}$ locations increase with increasing $R$.

**Accounting for "lost mass"**

The first step towards rationalizing the discrepancy in $N \mathrm{k} T = \frac{\rho \mathrm{R} T}{M_X}$, is the sol fraction estimate, provided by swelling experiments. As mentioned in Section 3, $f_{\mathrm{sol}}$ varies from ~0.11 at $R$=1 to ~0.41 at $R$=0.8 (Figure B-2). Thus, the gel fraction, $(1-f_{\mathrm{sol}})$, decreases from 0.89 to 0.59; i.e., by a factor of ~1.5.

We account for the remaining "lost mass" (where the increase in chain size, *n*, still does not account for all the decrease in the $N\,\mathrm{k}T$ ) in terms of the pendant chain fraction, $f_P$, of the total gel mass, which does not bear the load during deformation. Therefore,

$$N\,\mathrm{k}T = \frac{\rho \mathrm{R} T}{M_X}\left(1 - f_{\mathrm{sol}}\right)\left(1 - f_P\right) \tag{15}$$

For networks built from building blocks via step growth polymerization (such as PBPU elastomers and PDMS networks), pendant chains have been inferred via swelling experiments [27,47]. Significant pendant fraction polymers are also predicted via models such as the Flory-Marsh-α model [7,9,48,49] and the Miller-Macosko-Vallés approach [50–56]. Recently, stress-elongation data of near-critical sparsely crosslinked swollen gels, where the entanglement effects are suppressed, has been employed to infer pendant branches [57].

Eqn. (15) indicates that defects in the form of sol fraction and pendant chains, reduce the effective crosslink modulus. The defect fraction, $f_d$, then is

$$f_d = 1 - \left(1 - f_{\mathrm{sol}}\right)\left(1 - f_P\right) \tag{16}$$

Thus, the combination of eqns. (3) to (15), forms the basis for modelling the stress-elongation data, which provides insights into the variation of network morphology with the stoichiometry measure, *R*.

# 5 Inferring Network Architecture from Mechanical Behavior

## 5.1. MR Plot Modeling

There are two upturns in the MR plots, which the fitting models need to address. The two upturns originate from different phenomena, and it is necessary that the one upturn does not confound the understanding of the other.

The finite extensibility upturn at low $\frac{1}{\lambda}$, contains a high density of data points. This is because data points are collected at a uniform rate with respect to $\lambda$, and 0.16 < $\frac{1}{\lambda}$ <0.45 corresponds $\Delta\lambda \sim 4$, while 0.45 < $\frac{1}{\lambda}$ <0.9 corresponds to $\Delta\lambda \sim 1.1$. The relative magnitudes of $\Delta\frac{1}{\lambda}$ are reversed. The latter region is characterized by fluctuation suppression due to entanglements.

In order to rationalize the combined effects of two upturns – one in the high density data region and one in the wide expanse region, we carry out a preliminary fitting first – of the resultant model combining eqns. (3) to (15). The fitted parameters for this region, $f_P$, $n$, $G_e$, are not adequate to describe the entire data range.

For $\frac{1}{\lambda}$ >0.7 to 0.8, an upturn has been present in MR data reported since 1990 [1–4,12,18,19]. In data that span tension and compression, on an increasing $\log_{10}\frac{1}{\lambda}$ scale, this upturn at $\frac{1}{\lambda}$ ~0.7, transforms to a maximum at $\frac{1}{\lambda}$ ~1.8, via an inflection near the interphase between tension and compression ($\frac{1}{\lambda}$ ~1) [19]. Our data are limited to tensile extension. To the best of our knowledge, this upturn for low tensile deformation, has never been modeled. The extreme upturns in the very low deformation regions (both tension and compression) have been rejected as artefacts due to errors in estimating $\left(\lambda-\frac{1}{\lambda^2}\right)^{-1}$ [58].

However, acknowledging the consistent existence of upturn, leading to a maximum in the compression regime, indicates that our materials would exhibit an inflection – possibly at the cusp between the tension and compression regimes, near $\frac{1}{\lambda}$ ~1.0. The slip-tube model has been implemented over a limited deformation range on Gaussian networks [31]. Their comparison data [19–21,31] are relatively sparse in the low deformation regime. Therefore, their approximate fits did not find any upturn, when plotted against $\log_{10}\left(\frac{1}{\lambda}\right)$.

We assume that our experimental data and their combined tension-compression fits would exhibit clear inflections, about the maximum at $\log_{10}(1.8)$~0.25. Therefore, we add a phenomenological Gaussian factor to $\sigma_e$, to enable the model to reflect the experimental observations (an upturn on either side of the possible maximum at $\frac{1}{\lambda}$ ~1.8).

$$\sigma_{e,red}=\frac{G_e\left(1+G_e^G\,\mathrm{Exp}\left(-\frac{\left(\mathrm{Log}_{10}\left[\left(1/\lambda\right)\right]-0.25\right)^2}{2s^2}\right)\right)}{-0.35+0.61\left(\lambda\right)^{-0.5}+0.74\lambda} \tag{17}$$

Combining eqn. (11), eqn. (15) and eqn. (17),

$$\sigma_{red} = (1 - f_{sol})(1 - f_P)\frac{\rho \mathrm{R} T}{n M_S}\frac{2}{3n}\left(\left(\frac{n + 0.5n^{0.75}}{2} + \frac{(3 + 0.5n^{0.75} - 2n)}{2(1-\Lambda)}\right) - \left(\frac{3(\Lambda^2 - 6\Lambda + 3)}{(\Lambda - 3)^2}\right)\right)$$
$$+ \frac{G_e\left(1 + G_e^G \mathrm{Exp}\left(-\frac{(\mathrm{Log}_{10}[(1/\lambda)] - 0.25)^2}{2s^2}\right)\right)}{-0.35 + 0.61(\lambda)^{-0.5} + 0.74\lambda} \quad (18)$$

The $s$ inside the Gaussian exponential of eqn. (17) is its standard deviation, i.e., the distance of the inflection from the maximum at ~0.25, on the $\log_{10}\left(\frac{1}{\lambda}\right)$ axis.

With this basis, we consider the MR plots of $\sigma_{red}$ vs $\frac{1}{\lambda}$, for the 5 stoichiometries, *R*=1.0 to *R*=0.8, and fit eqn. (18) (Figure 6). The fitted parameters are $f_P$, $n$, $G_e$, $G_e^G$ and $s$. Employing 5 fitted parameters is *prima facie*, a limitation of the modeling method. However, of these parameters, $f_P$, $n$, $G_e$, would similar to those estimated via eqns. (3) to (15).

Figure 6 is a representative depiction of the performance of modeling via eqn. (18). In addition to the excellent visual correspondence the minima are locations are also reproduced by the model (<1% relative error in the $\frac{1}{\lambda}$ value). The resultant model parameters are listed below in Table 1. The relative standard deviation <5% for all parameters. We next interpret the modeling results in terms of the network morphology.

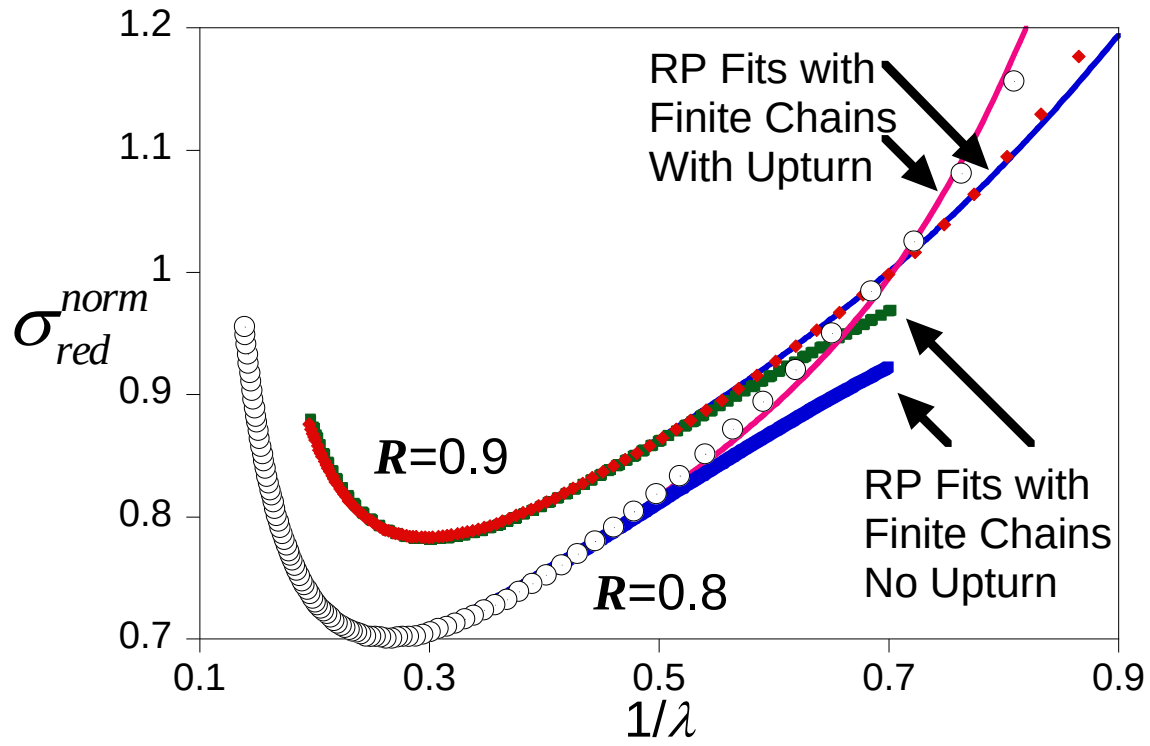

**Figure 6.** 3-Parameter fits (no-upturn) (eqns. (3) to (15)) and eqn. (18) fits (for the upturn)

Table 1. Best fit parameters for eqn.(18)

| $R$ | $f_P$ | $n$ | $G_e$ | $G_e^G$ | $s$ |
|---|---|---|---|---|---|
| 1.00 | 0.248 | 28.3 | 0.328 | 1.0 | 0.154 |
| 0.95 | 0.39 | 32.6 | 0.226 | 2.2 | 0.123 |
| 0.90 | 0.368 | 52.1 | 0.105 | 1.0 | 0.170 |
| 0.85 | 0.411 | 60.0 | 0.0706 | 2.0 | 0.168 |
| 0.80 | 0.663 | 75.3 | 0.0284 | 3.9 | 0.148 |

### 5.2. Molecular interpretation of Modeling Results

We examine first the crosslink modulus, $G_X$ (combining eqns. (5) and (15)). The fitting equation of Figure 7, indicates one effect of *R* on the network morphology. We find that when $G_X = 0$ (non-percolation of crosslinks), *R*~0.76. This is consistent with the observation that the network does not percolate in the aborted (no reportable data) *R*=0.75. Therefore, *R*=0.8 could be a transition between a well percolated network and non-percolating crosslinks.

As *R* decreases, the effective defect fraction, $f_d$ (eqn. (16)), increases, causing $G_X$ to decrease. Figure 8 plots the gel fraction $\left(f_{\text{gel}} = 1 - f_{\text{sol}}\right)$, the main-chain fraction $\left(f_{chain} = 1 - f_P\right)$ and the overall network fraction $\left(f_{\text{net}} = 1 - f_d\right)$.

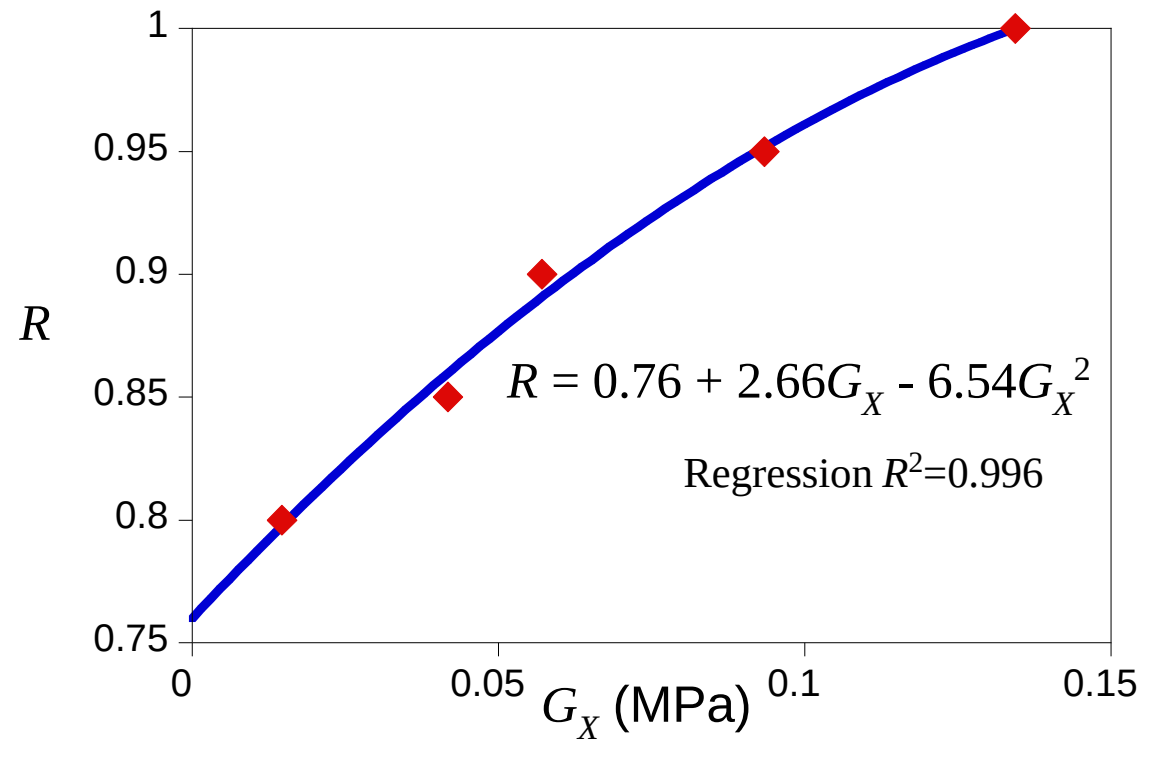

**Figure 7.** Percolation limit for stoichiometry, via crosslink modulus. When *R*~0.76, $G_X$~0,

The swelling experiment $f_{\text{sol}}$ data and the model parameters, $n$ and $f_P$ indicate a morphology for PBPU networks, as illustrated in Figure 9. The schematic of the building blocks of the networks – network chain with branch(es) and sol chains – for *R*=1.0 and *R*=0.8, are illustrated in Figure 10. Appendix C describes the rational application of the Marsh-α model to explore this aspect.

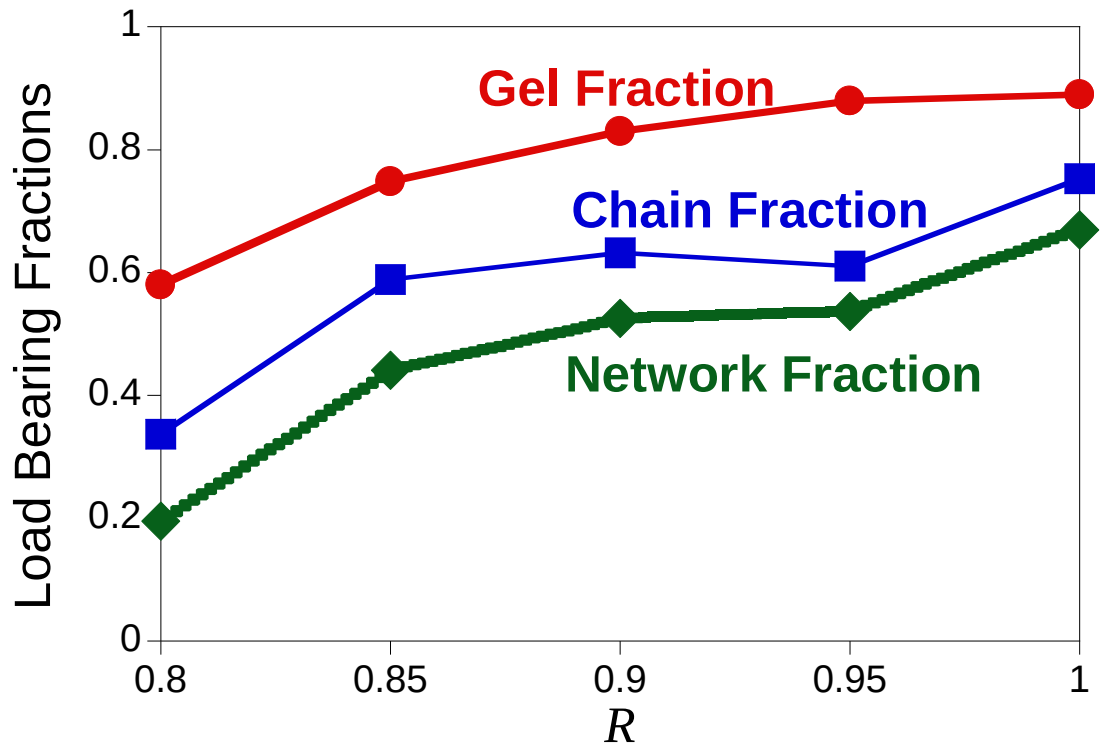

**Figure 8.** Load Bearing fractions of the network – gel vs sol, main-chain vs pendant, load-bearing network vs defect, as function of *R*. The connecting lines are to aid the eye.

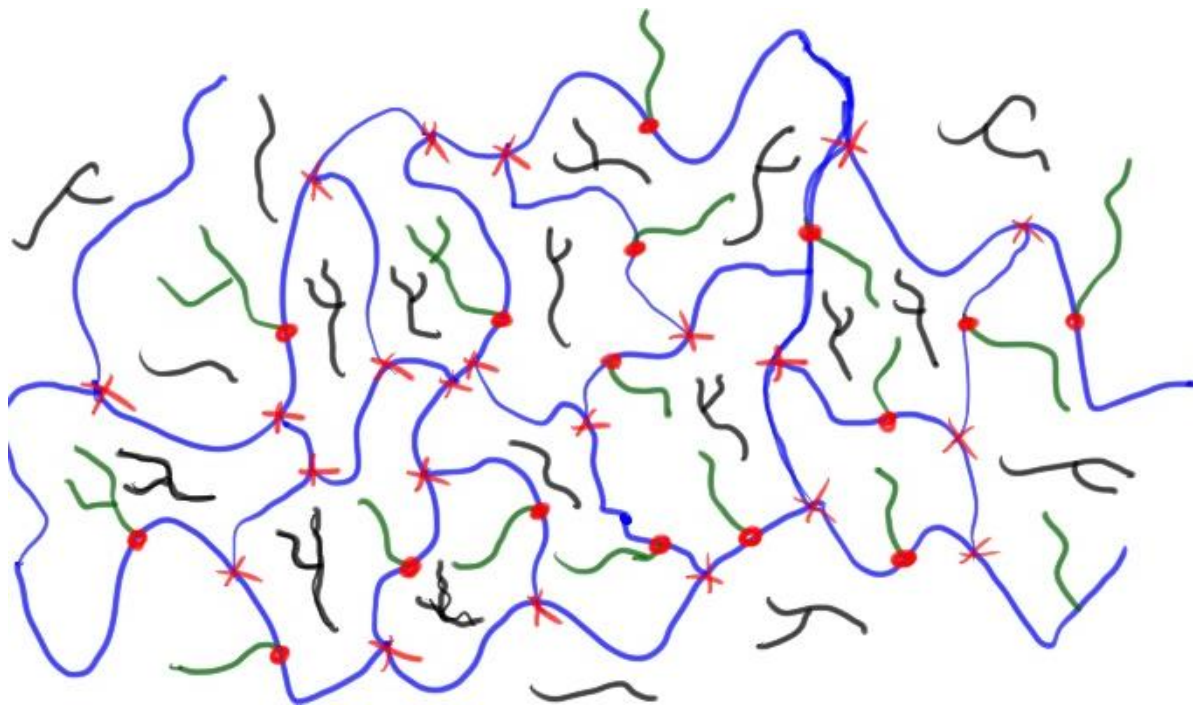

**Figure 9.** PBPU elastomer network bulk comprising trifunctional junctions as red crosses (color online) for blue network chains and as red filled circles for green branching chains. The free-floating linear and branched black chains are the sol component.

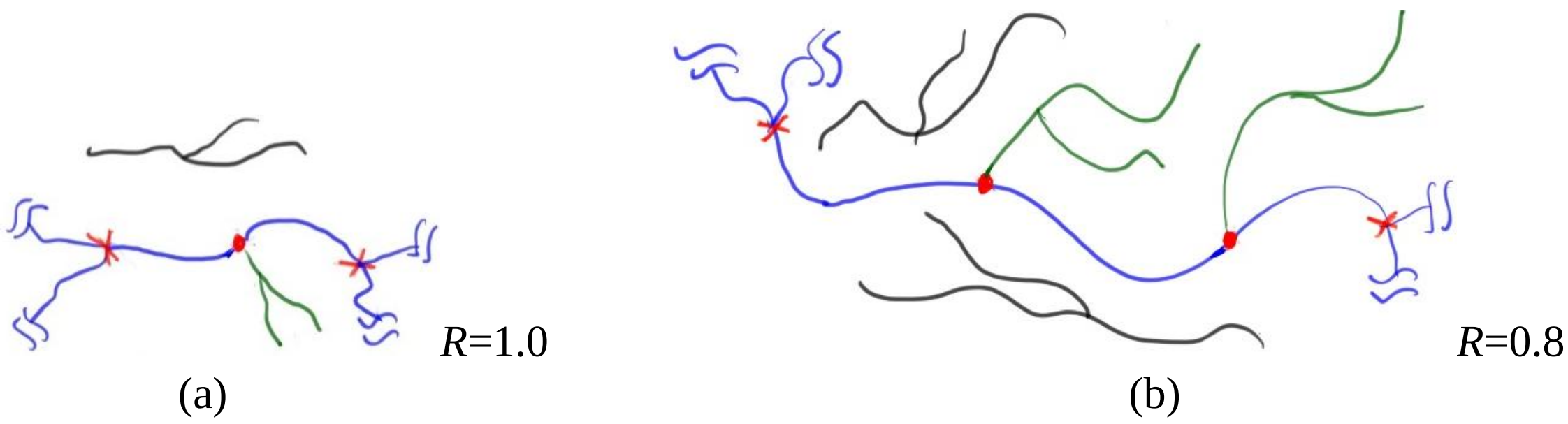

**Figure 10.** PBPU elastomer bulk building blocks Schematics. Entity color same as in Figure 9: (a) $R$=1.0, $f_P$ ~0.25, $n$ ~28, $f_{\text{sol}}$ ~ 0.11 (b) $R$=0.8, $f_P$ ~0.66, $n$ ~ 758, $f_{\text{sol}}$ ~ 0.41.

This schematic is consistent with the failure of samples of *R*=0.95 and *R*=1.0 before reaching the minima in their MR plots. Heterogeneity in the network will give rise to some chains of shorter lengths, resulting in early failure [59,60]. Heterogeneity effects in lower *R* samples might be mitigated by the plasticization effects due to their greater $f_{\text{sol}}$ and $f_P$ .

Based on the above network descriptors (chain size and load bearing fractions), we examine other network features. Figure 11 depicts the effect of stoichiometry on the crosslink modulus, $G_X$, the MR modulus, $G_e + G_X$, and the initial modulus, $G_e^{\max} + G_X$.

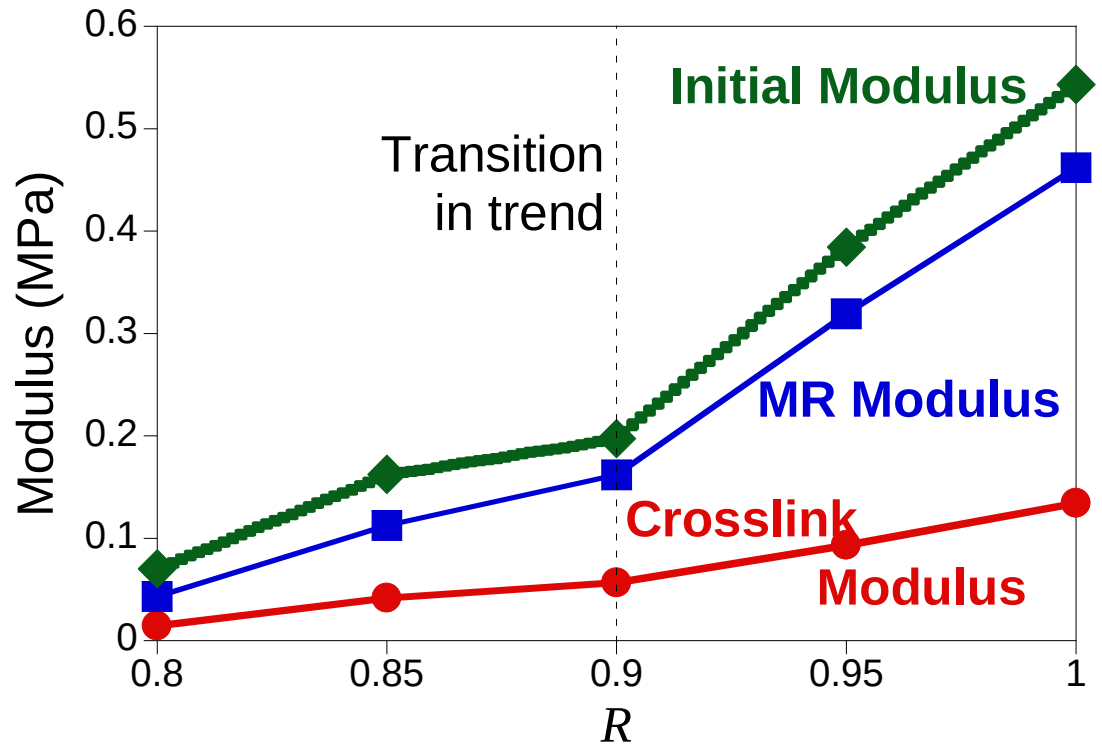

**Figure 11.** Crosslink Modulus, Effective MR Modulus and Initial Modulus as function of *R*. There appears a change in the trend (slope) at *R*=0.9. Data connected to aid the eye.

Figure 12 delves into various features of the entanglement modulus.

The ratio, $G_e/G_X$ lies in the range 1.7 to 2.4. For elastomers formed by curing (crosslinking) of long-chain molecules, this ratio is usually <1. The greater value indicates the increased severity of the entanglement effects due to the presence of branches on the entangled chains. The effect of these branches is highlighted by the significant increase in $G_e^{\max}/G_e$ with increasing branching fraction (decreasing *R*).

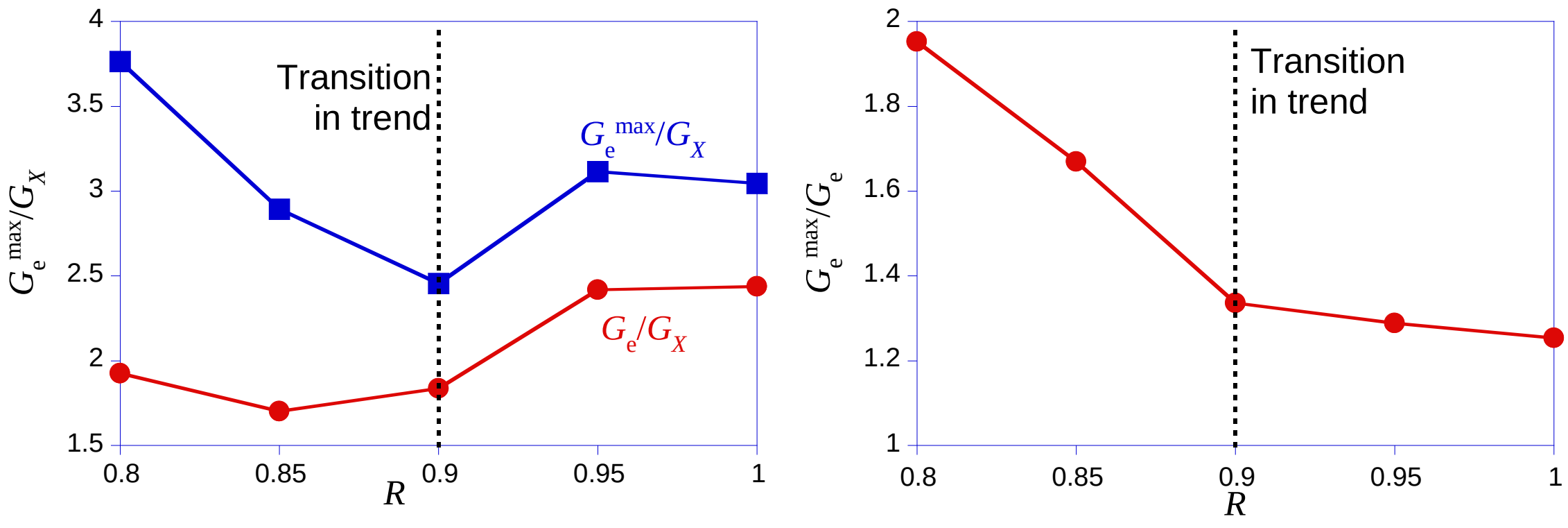

**Figure 12.** Normalized Entanglement effects vs *R* (a) $G_e^{\max}/G_X$ and $G_e/G_X$ (b) $G_e^{\max}/G_e$. The connecting lines are to aid the eye

Next, we consider the effect of the network morphology on the swelling behavior. We map the known values of $M_X$ to eqn. ( 20), to estimate the Flory-Huggins $\chi$ parameter as function of *R*. Our experimental range (Figure 13) corresponds well with the reported values. $\chi$~0.36 [8,23,61], has been estimated with the solubility parameter, $\delta_{HTPB} \sim 17.5\left(\text{J/cc}\right)^{0.5}$,via group contribution methods $\left(\delta_{Toluene} \sim 18.1\left(\text{J/cc}\right)^{0.5}\right)$.

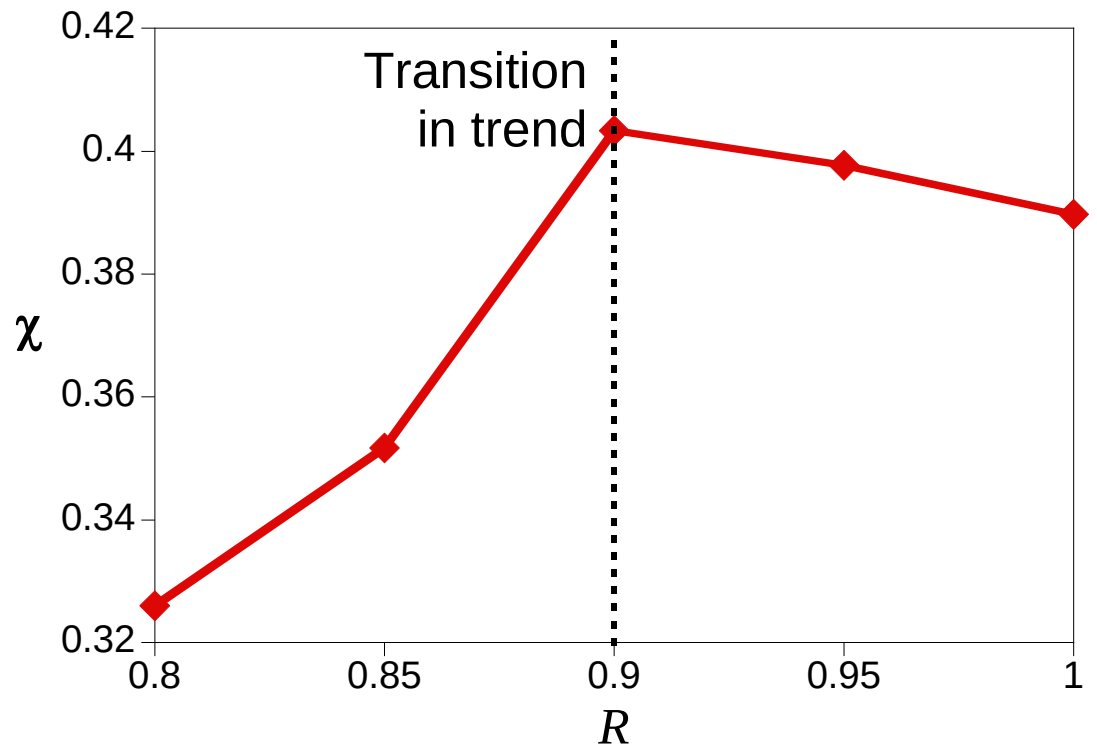

**Figure 13.** Flory-Huggins $\chi$ parameter as function of *R*.

In Figure 11 to Figure 13, we find a qualitative transition at *R*~0.9, where there appears to a plateau of all moduli for. At low *R*, the "relative entanglement modulus" is lower, but the low deformation entanglement effects due to the pendant chains is greater. We also find greater values ($\chi$~0.4), when the short-branch chains HTPB density decreases up to a point, as the branch length increases (from *R*=1.0 to *R*=0.9); then $\delta_{HTPB}$ also decreases, and increases $\chi$. Further increase in branch length increases HTPB density, thus increasing $\delta_{HTPB}$, and decreasing $\chi$ (the swollen networks have released the sol components).

Finally, we verify that the overlap mass concentration (0.022g/cc) of an FJC of 75 segments of length *a*~11.1 Å, $M_s \sim 135$, is less than $\rho \times \left(1 - f_d\right)$=0.186g/cc, which indicates that the network chains themselves are well entangled with each other; the chains have their unperturbed sizes when the network bulk is in the undeformed state.

# 6 Conclusions

We have examined the mechanical behavior polybutadiene polyurethane (PBPU) elastomers formed with stoichiometric ratio *R*=[NCO]/[OH] varying from 0.8 to 1.0. Our experimental methods and their subsequent analyses have revealed that the reaction scheme leads a crosslinked network with *R*-dependent pendant fraction.

The experimental procedure involves deformation onset errors due to approximations in the gauge length and preload. There are also errors due slippage of the laser extensometer reflector strips. We mitigate these errors by comparing the extensometer measurement with the sample-holder crosshead displacement, to estimate the correct stress-elongation data.

For the chemically crosslinked network, we have employed the phantom network deformation model for ideal freely jointed chains, where the junction fluctuations that relax part of the stress, are themselves suppressed during deformation due to finite chain effects. We superimpose the entanglement part of the slip-tube model, because its entanglement component is distinct from the crosslink component. When combined with the finite chain effects, this model becomes convex upward at $\frac{1}{\lambda} > 0.5$. However, all data since 1990, including those for combined tension-compression experiments exhibit upward concave behavior in the region $0.7 < \frac{1}{\lambda} < 0.9$, not accounted for by any model. Our approach addresses this shortcoming.

We implement corresponding deformation state analyses of the Mooney-Rivlin data by normalizing $\sigma_{red}$ data for each MR data set its own reference state, $\sigma_{red}\big|_{\frac{1}{\lambda}=0.7}$. The intrinsic curvatures of the plots, become visible on this self-scale. They indicate that the number of segments constituting network chains is much lower than that corresponding to the $\sigma_{red}$ magnitudes. This discovery, along with the sol fraction from swelling experiments, reveals that the "lost mass" is located as pendant branches on the load-bearing network chains. Thus corresponding states normalization overcomes the camouflaging effect of the necessary 5 parameter accurate modeling of the MR data to reveal the network morphology.

# Appendices

## A. Estimation of Rectified Stress-Elongation Data

There are a few concerns with this established protocol. These are listed and addressed here.

1. The expectation with the preload is that it will only straighten the sample and make it taut, without stretching it. However, for a sample cross section of 6mm×3mm, the preload corresponds to a stress of 5.5 kPa. The PBPU elastomers in this study, undergo 2%-5% extension, over this stress range. Hence, subtracting the preload is incorrect, because this force is present in the material, throughout the experiment [14]. This force causes an extension which is lost in axes-shifts. It is necessary to accurately estimate this unaccounted for extension, and add it suitably to the measured extension.
2. Since the silver strips are stuck and wrapped manually, there is uncertainty in the accuracy of the stated gauge length, and thus, in the corresponding strain.
3. We find that in the region close to $\lambda$~1, there is a significant scatter – since slight errors in determining the extension of the sample, cause significant errors in $\sigma_{red}$.

The justification for rectifying the apparently small errors, particularly in the low stress regions (even though they are far removed from the performance regions), is in their effect on the estimates of the network structural parameters. The limiting slope and intercept at $\lambda$=1, provide quantitative measures of the crosslink modulus and the entanglement constraint modulus. Therefore, it is necessary to correctly estimate the experimental data near $\lambda$~1, in order to correctly estimate the model parameters, to understand the physical phenomenon better.

These uncertainties and concerns, are addressed next. We describe first, the method to account for this preload extension and employ this correction to determine the MR plot. We consider the typical data provided by the tensile tester (Table A-1); these are recorded only after the ~0.1 N preload.

We linearly extrapolate the cross-head data up to 0 N force, to identify its location in the absence of force. In this particular example, that location is at 0.027 mm. Hence, the force vs crosshead plot should have its origin shifted accordingly. The crosshead position values are now reduced by this amount, and we now consider the new values, $X$.

**Table A-1.** Typical initial data provided by the UTM. Actual data are recorded until failure (~600% strain)

| Force (N) | Cross-Head (mm) | Strain (%) | Time (min) |
|---|---|---|---|
| 0.09 | 1.26 | 0 | 0.0316 |
| 0.117 | 1.63 | 0.563 | 0.0394 |
| 0.14 | 2.02 | 1.28 | 0.0472 |
| 0.154 | 2.41 | 2.14 | 0.055 |

Similarly, the strain data are based on the input gauge length value of 25 mm. Therefore, only strip distances greater than 25 mm are considered as strained. We can obtain the sample extension, $l_e$, from the strain values provided by the UTM and the extensometer. The $l_e$ values in mm for rows 2, 3, and 4 are 0.1475, 0.32, 0.535, respectively. These are the distances between the silver strips, beyond the input gauge length, 25 mm. The actual strain is unknown, because the actual gauge length is unknown.

In order to estimate the real gauge length, we consider the plot of the ratio $l_e/X$ vs $X$. The plot for this sample is provided in Figure A-1.

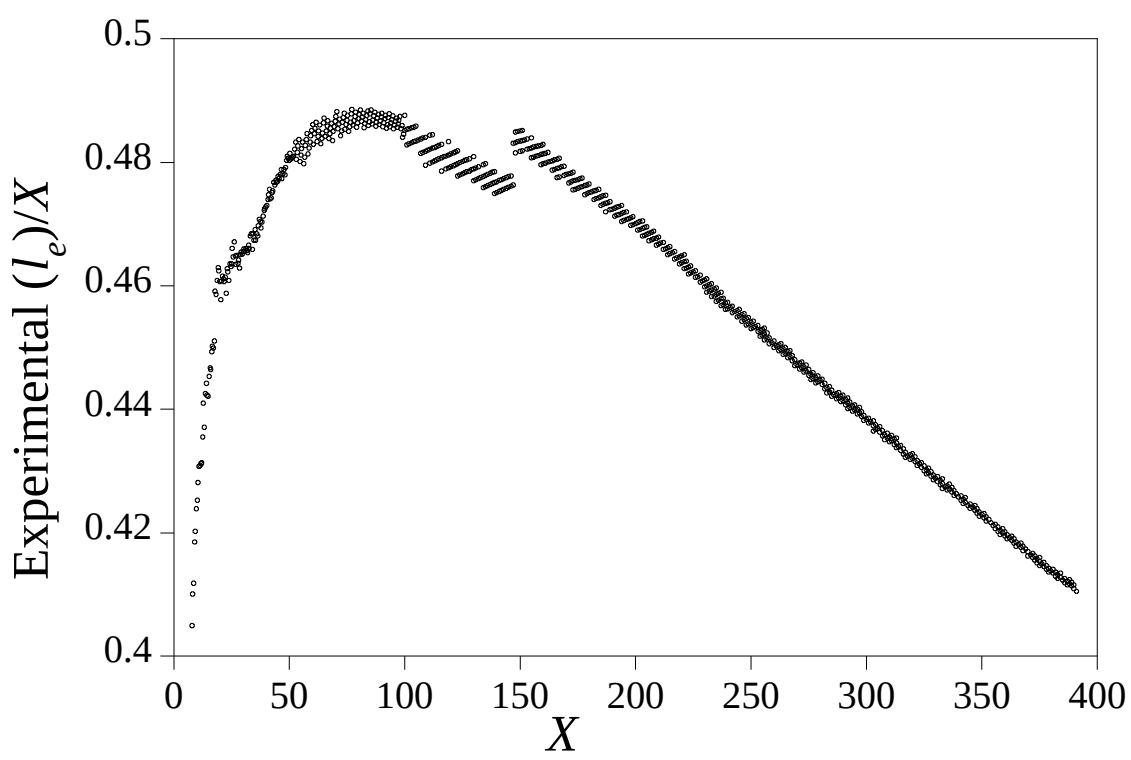

**Figure A-1.** Ratio of sample extension to crosshead movement (relative to extrapolated zero load values)

In Figure A-1, the estimated $l_e/X$ initially increases with $X$. The ratio is very small for small $X$. This is because the error caused by the assumed gauge length is greater; e.g., for the first point at $X$=1.63 mm, and $l_e$~0.1475 mm, and distance between the silver strips being 25.1475 mm; if the actual gauge length were 24.8 mm, then $l_e$~0.3475 mm, and the strain would be 1.4%, instead of the provided 0.563%. The error in estimating the actual $l_e$ decreases as $X$ increases; i.e., as the artefacts due to the incorrect gauge length become less dominant. Figure A-1 indicates that $l_e/X > 0.4$, only after the crosshead rises 7.5 mm. The rise is steep, and there is a discontinuity in the slope at $X$~30 mm. Subsequently, beyond $X$=90 mm, $l_e/X$ decreases with $X$, up to $X$~140 mm.

Over the range from $X$=140 mm to $X$=150 mm, we see an upward jump in $l_e/X$ from ~0.475 to ~0.485. The discontinuity in the data occurs at $X$=147 mm. This jump is very likely due to the slippage of the lower silver strip (Figure A-2). We address such issues in our analyses of the data, recognizing that the crosshead control is precise, as is the reading of the force transducer.

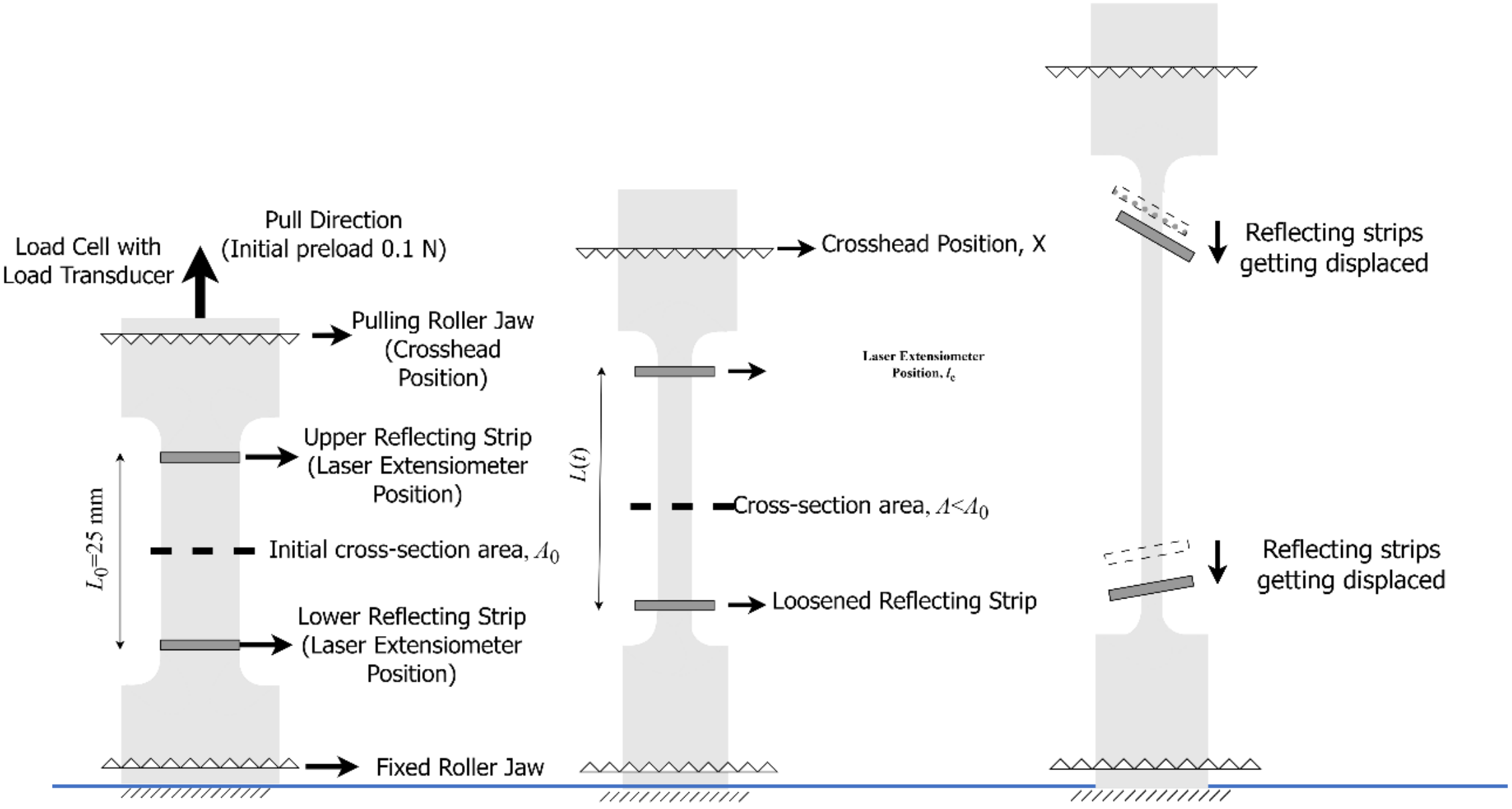

**Figure A-2.** The deforming sample indicating the grips, crosshead, extension, cross sectional area and the laser extensometer reflector silver strips, which may slip during deformation.

The objective is translating these inferences into providing accurate stress-elongation data. We carry out the following steps:

1. We consider first, the region around $X$=80 mm, where $l_e/X$ exhibits a maximum. We fit the region from $X$=50 mm to $X$=90 mm, to a quadratic, and locate the maximum ($l_e/X$~0.487) at $X$~84 mm.

2. Next, about the maximum, we consider individual LHS and RHS ranges: the LHS range from $X$=40 mm (beyond the discontinuity at $X$~30 mm) to the maximum at $X$~84 mm, and the RHS range, from the maximum at $X$~84 mm up to the discontinuity at $X$=147 mm. In Figure A-3, these are depicted in blue for the LHS and in red for the RHS. The individual fit of each parabola, $\left(l_e/X\right)_{\max} - \left(l_e/X\right) = a_i \times \left(X - X_{\max}\right)^2$, ($a_i = a_L, a_R$, for the LHS and RHS curves respectively) yields $a_L \sim 6.52\times10^{-6}$, $a_R \sim 3.56\times10^{-6}$. We assume that by $X$~40mm, the artefacts due to the incorrect gauge length (error ~0.2 mm), are negligible. These ranges are determined visually, assuming that $l_e/X$ follows a parabolic relationship with $X$. Then the ratio, $l_e/X$, would vary only because of the actual deformation, vis-à-vis the crosshead separation.

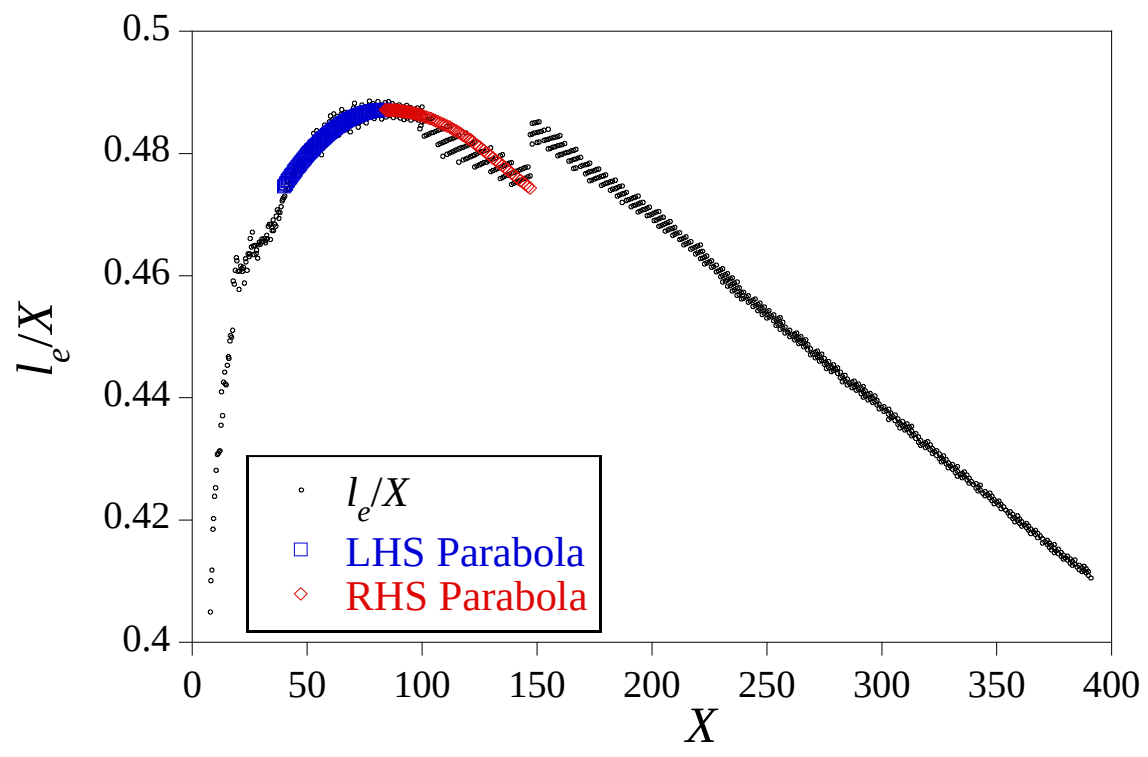

**Figure A-3.** Parabolic smoothening of $l_e/X$, about its maximum.

3. For $|X_{max} - X| >\sim 40\,\text{mm}$, we assume that the variation is linear and continuous with that at $|X_{max} - X| \sim 40\,\text{mm}$, depicted as a red line in Figure A-4. We implement this idea directly on the LHS, by obtaining the slope at $|X_{max} - X| \sim 40\,\text{mm}$. This slope $\sim 80 \times a_L \sim 5.36 \times 10^{-4}$.

4. On the RHS, we negate the upward jump at the discontinuity, which is an experimental artefact. We linearly fit the data (jumped up data in Figure A-3, beyond the RHS parabola) from the discontinuity at $X$=147 mm, up to the maximum crosshead location, $X$=390mm, to find $\left(l_e/X\right) = -3.08 \times 10^{-4} X + 0.53$.

5. Thus, we locate the point, $X$~127 mm, on the RHS parabola, where slope $\sim -3.08 \times 10^{-4}$. Then, we consider that $l_e/X$ varies linearly with $X$, beyond $X$~127 mm. After complete alignment of the parabolic data, we correct the data 127 mm<$X$<147 mm (green line).

6. For continuity in $l_e/X$ at 147 mm, the data beyond $X$=147 mm, need to be shifted downwards by 0.0113; the $l_e/X$ values from $X$~127mm to $X$=147 mm, would then vary linearly via slope $-3.08 \times 10^{-4}$, so that they are collinear with the data for $X$>147 mm. The RHS plot is shown in Figure A-5.

7. Combining LHS and RHS, the overall plot is given below (Figure A-6).

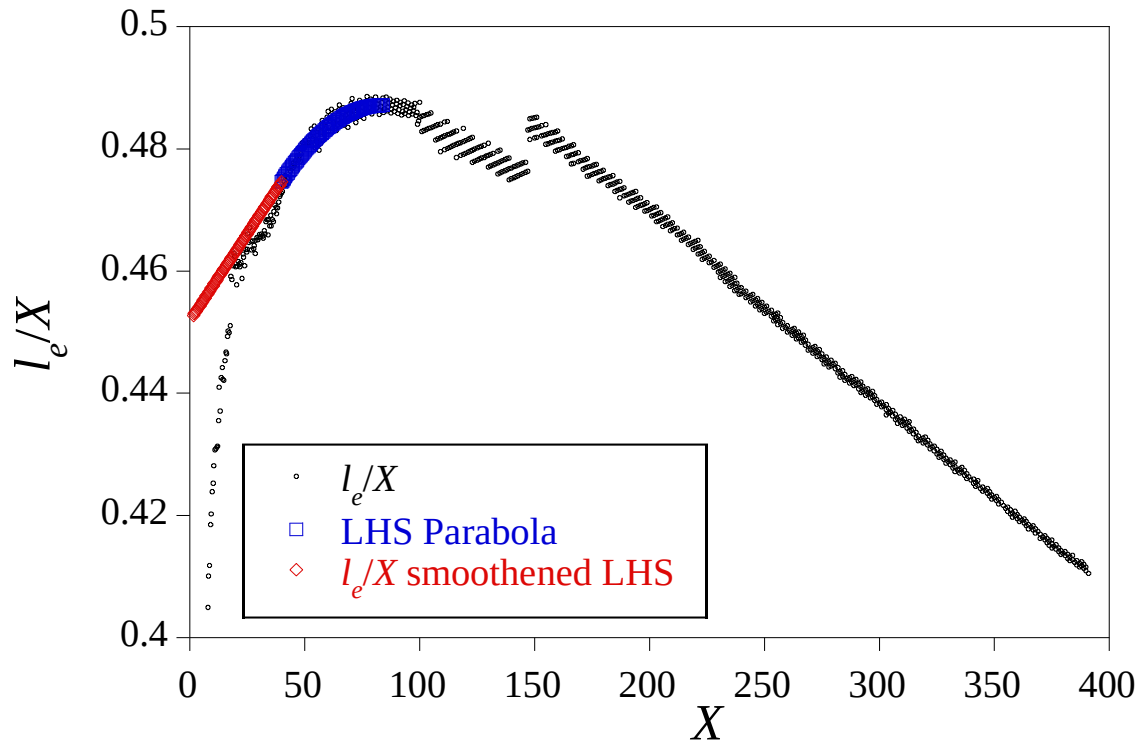

**Figure A-4.** Locating the zero-$X$ intercept of $l_e/X$.

8. The next step is determining the actual extension of the sample, and converting this extension to the strain. This extension, $l_e = (l_e/X) \times X$ , is directly obtained by multiplying the coordinates of the points on the final curve in Figure A-6.

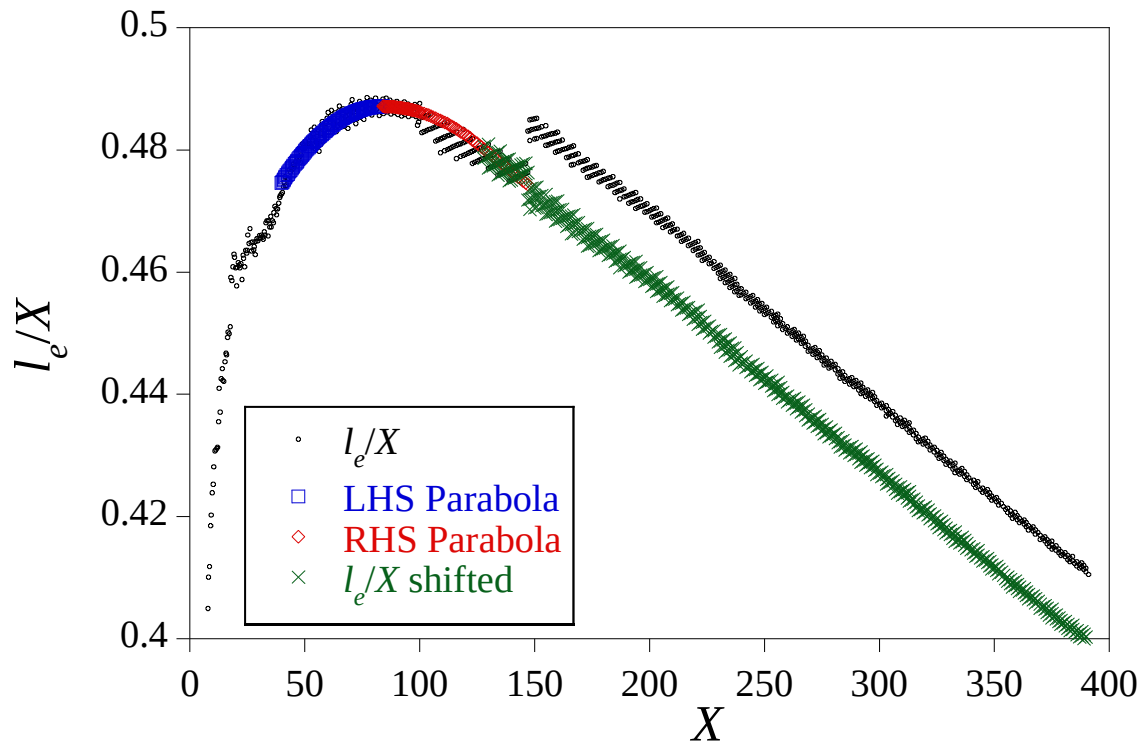

**Figure A-5.** Accounting for the jump in $l_e/X$, possibly due to shifting of the reflective strips

9. In order to obtain the strain from extension estimates, it is necessary to determine the actual gauge length. The input gauge length, $L_0^{in}$ , at the preload is 25 mm. We recognize that the extensometer determines the instantaneous distance between the silver strips and computes the strain, based on the input. Hence, the deformed length, based on the extensometer, i.e., $\left(L_0^{in} + l_e^{ex}\right)$ equals the deformed length based on the actual gauge length, i.e., $\left(L_0 + l_e\right)$.

   Therefore, the actual gauge length is obtained from $\left(L_0 + l_e\right) = \left(L_0^{in} + l_e^{ex}\right)$.

   The value of $L_0$ , at $X$=0, corresponds to the actual sample gauge length.

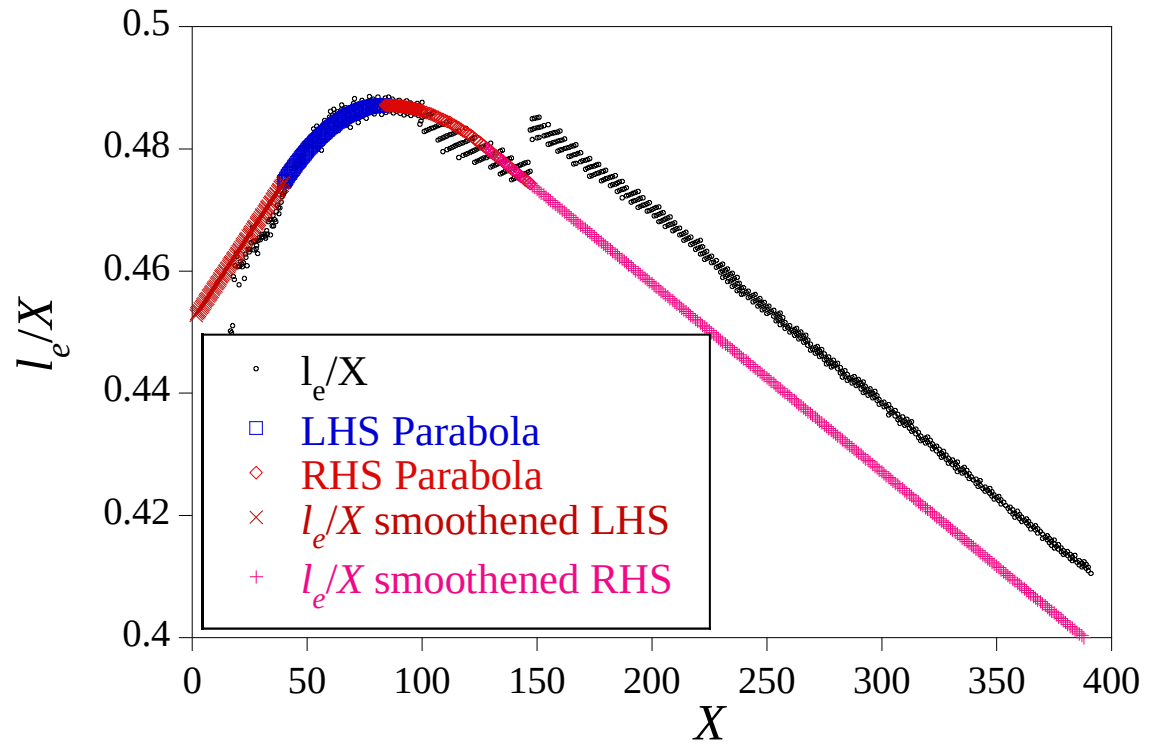

**Figure A-6.** Linear smoothening of the mid to high elongation data

The actual strain, $\varepsilon = {l_e}/{L_0}$, and the elongation ratio, $\lambda = 1 + \varepsilon$.

# B. Equilibrium Swelling Experiments

1. Minimum swelling times were determined for achieving equilibrium through swelling kinetic studies (Figure B-1). The samples were maintained under dark conditions with no exposure to air to reduce the degradation effects.

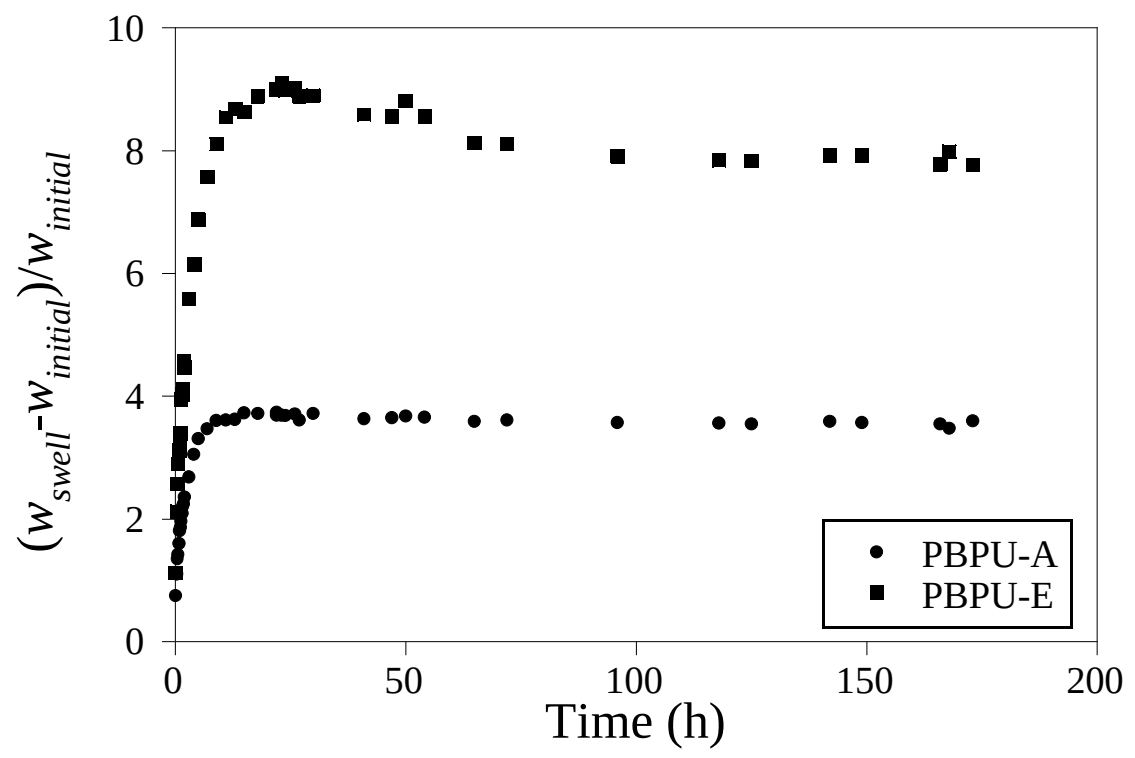

**Figure B-1.** Degree of swelling for samples with highest (PBPU-A) and lowest (PBPU-E) R-values.

2. Small cubes weighing approximately 0.10 to 0.15 g were cut, and the samples' initial weights were recorded. Three sets of experiments were carried out for each R-value.
3. All the samples were kept in 50 ml toluene in different beakers for 72 hours. Subsequently, the toluene was changed, and samples were kept for a further 72 hours.
4. The samples were then removed from toluene and gently wiped, and the weights of swollen samples were recorded immediately $(w_s)$.
5. The samples were dried at 105°C for 3 hours with intermittent vacuum application at 2 torr.
6. Final weights of samples were recorded after de-swelling $(w_{ds})$.

7. All weighments were carried out with a balance having an accuracy of $10^{-5}$ g.

The classical Flory-Rehner equation is widely used for determining the lengths of effective chains in polymer networks. This can be via the affine network model (eqn. ( 19)),

$$M_X = \frac{\rho V_s \left( \upsilon_2^{1/3} - \left(1 - \frac{2}{f}\right) \upsilon_2 \right)}{\ln\left(1-\upsilon_2\right) + \upsilon_2 + \chi \upsilon_2^2} \tag{19}$$

or the phantom network model (eqn. ( 20)).

$$M_X = \frac{\rho V_s \upsilon_2^{1/3} \left(1 - \frac{2}{f}\right)}{\ln\left(1-\upsilon_2\right) + \upsilon_2 + \chi \upsilon_2^2} \tag{20}$$

$\upsilon_2$ is the volume fraction of the polymer, $M_X$ is the molecular weight of the effective chains between crosslinks, $\chi$ is the Flory-Huggins polymer-solvent interaction parameter. In this method, $\upsilon_2$, in the swollen sample is experimentally determined as described above.

$$\upsilon_2 = \frac{w_2 / \rho_2}{w_1 / \rho_1 + w_2 / \rho_2} \tag{21}$$

where, $w_1$ and $w_2$ are the weight fractions of solvent and polymer, respectively, in the swollen sample. $\rho_1$ and $\rho_2$ are the densities of the solvent and polymer, respectively. The weight fractions can be obtained from the swelling ratio, *Q*.

$$w_2 = \frac{1}{1+Q} \tag{22}$$

$$w_1 = 1 - w_2 \tag{23}$$

$$Q = \frac{w_s}{w_{ds}} - 1 \tag{24}$$

$w_s$ and $w_{ds}$ are the weights of swollen and de-swollen samples, respectively. The deswollen sample has lost its sol component, from its original dry state, prior to swelling (see Figure B-2). Determination of $\chi$ represents a significant challenge in polymer chemistry even today.

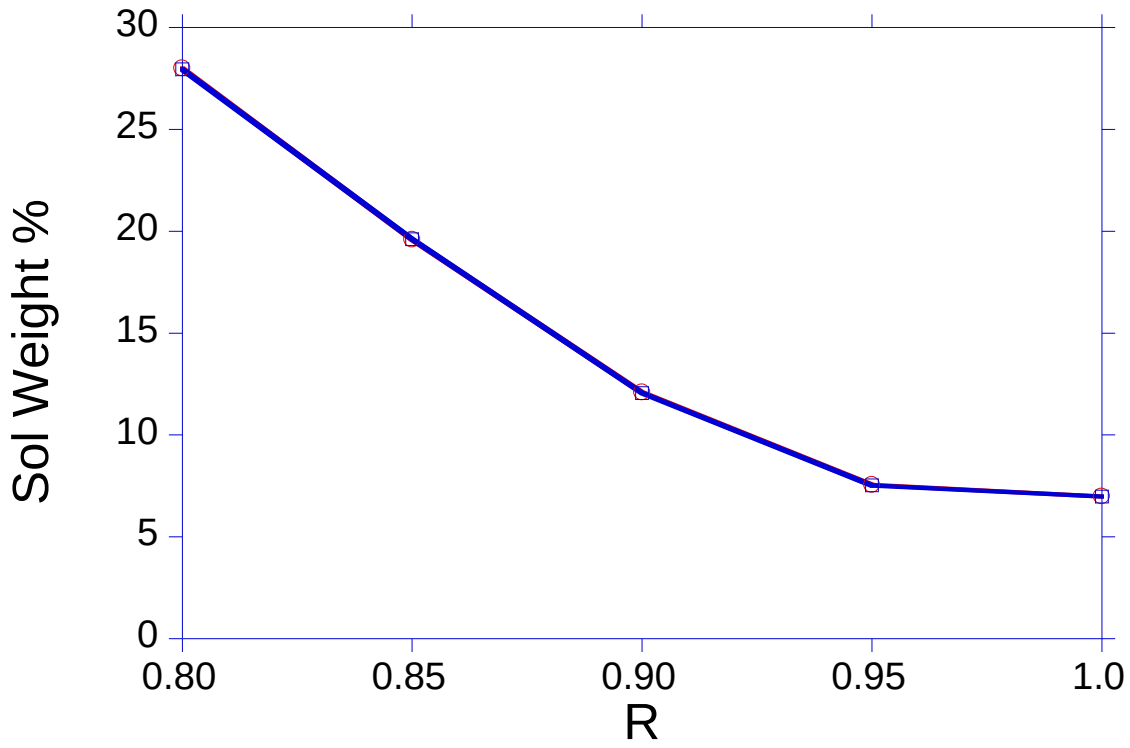

**Figure B-2.** Sol weight percent as function of stoichiometry.

## C. Marsh-α model framework

The following equations evaluate the network parameters for a polyol (component A with *OH* functionality) and a difunctional curing agent (component B with *NCO* functionality). We follow the notation of the original development by Marsh [48,49,62].

$$R = \frac{[B]}{[A]} = \frac{[p_A]}{[p_B]} \qquad (25)$$

$$\alpha = \frac{p_A p_B A_3}{1 - p_A p_B A_2} \qquad (26)$$

$$W_s = \left(\frac{1-\alpha}{\alpha}\right)^3 \qquad (27)$$

$$\nu_{Marsh} = \left(\frac{2\alpha - 1}{\alpha}\right)^3 \left(\frac{\rho A_3}{2W_{eq}}\right) 10^6 \qquad (28)$$

$$L_0 = \frac{W_g W_{eq} L_{sp} \alpha^3}{A_3 (2\alpha - 1)} \qquad (29)$$

$$R_m = \frac{1-\alpha}{\alpha} \qquad (30)$$

$$X_m = \frac{1}{R_m + 1/R_m} \qquad (31)$$

$$R_{m+1} = (R_m)^2 \qquad (32)$$

$$L_m = L_0 (1 + 2X_1)(1 + 2X_2)\ldots(1 + 2X_m) \qquad (33)$$

$$L = L_m \text{ as } X_m \to 0 \quad (34)$$

$$L_X = 2L \quad (35)$$

Here, $\alpha$ is the branching coefficient, [$A$] and [$B$] are the gram equivalent of polyol and curing agent, $A_3$ is the mole fraction of hydroxyls on tri-functional components of $A$, $A_2$ is the mole fraction of hydroxyls on di-functional components of $A$, $p_i$ is the fraction of the component $i$ reacted, $W$ is the weight fraction of sol, $W_g = 1 - W$, is the weight fraction of the gel, $L_{sp}$ is the number of chain atoms per gram (MW=54 for butadiene monomer, containing 4 chain atoms), $W_{eq}$ is the equivalent weight of the curing system, $\rho$ is the density of the polyurethane (g/cm$^3$), HCG is the moles of half chain per equivalent of B, $\nu_{Marsh}$ is the crosslink density (moles/m$^3$), $X_n$ is the chain extension coefficient, $L_0$ is chain atoms per 0-order half chain, and $L$ is the effective half-chain length (number of main-chain atoms). $L_X = 2L$, is the effective chain length between crosslinks.

The method for determination of trifunctional content assumes tri-functionality of the *G*-content in HTPB [7–10], which has been contested [63–66]. We assume trifunctionality to be based on the *G*-content, because extensive studies [7,67,68] on HTPB have indicated so, although this match may have been fortuitous. This could be one of the limitations of the Marsh method.

The Marsh method also provides for the determination of the distribution of pendant chains or dangling chains lengths ( $L_{P,m}$, for chains of order "*m*"). These chains are not necessarily linear chains, but will themselves contain branches and sub-branches. However, for order 1,

$$L_{P,1} = 2L_0 \quad (36)$$

For order 2 and above.

$$L_{P,m} = 2L_{m-1} + L_{P,m-1} \quad (37)$$

The average number of *m*-order pendent chains per effective network chain (see Figure B-2 for the distribution),

$$P_{m/X} = 2X_m \quad (38)$$

$$P_{m/X} = 2X_{m-1}\left(1 + 2X_m\right) \quad (39)$$

….

$$P_{1/X} = 2X_1(1+2X_2)...(1+2X_m) \quad (40)$$

To the above set of published equations, we add resulting equations, relevant to this work. The total number of pendant chains, per load-bearing chain:

$$\frac{N_P}{N} = \sum_{m=1}^{\infty} P_{m/X} \quad (41)$$

The total mass of all pendant chains on each load-bearing chain is:

$$M_P^{Tot} = \frac{1}{L_{sp}} \sum_{m=1}^{\infty} L_{P,m} P_{m/X} \quad (42)$$

The pendant chain fraction, then is

$$f_P = \frac{M_P^{Tot}}{M_P^{Tot} + M_X} \quad (43)$$

And the total MW between crosslinks would become

$$M_X^{Tot} = M_P^{Tot} + M_X \quad (44)$$

We first estimate *p* by comparing equation ( 35) (Figure C-1), with $M_X$ from stress-elongation modeling. The $M_X$ corresponding to the phantom network is too low, and will lead to limiting reagent extent of reaction, $p_B > 1$, for all *R* values. Hence, we select the applicable $M_X$ via an indirect process. We first consider the Flory-Huggins $\chi$ parameter as estimated via eqn. ( 20) (see Figure 13) for the phantom network in swollen elastomers. We then recompute $M_X$ via eqn ( 19), for an affine network [7–10].

Based on this $M_X$, mapped to equation ( 35), we compute the pendant fraction via equations ( 36) to ( 43). Figure C-2(a) depicts the distributions of pendant chain lengths. Adding all the pendant chains via eqn. ( 41), yields the number of pendant chains per load-bearing network chain (Figure C-2 (b)).

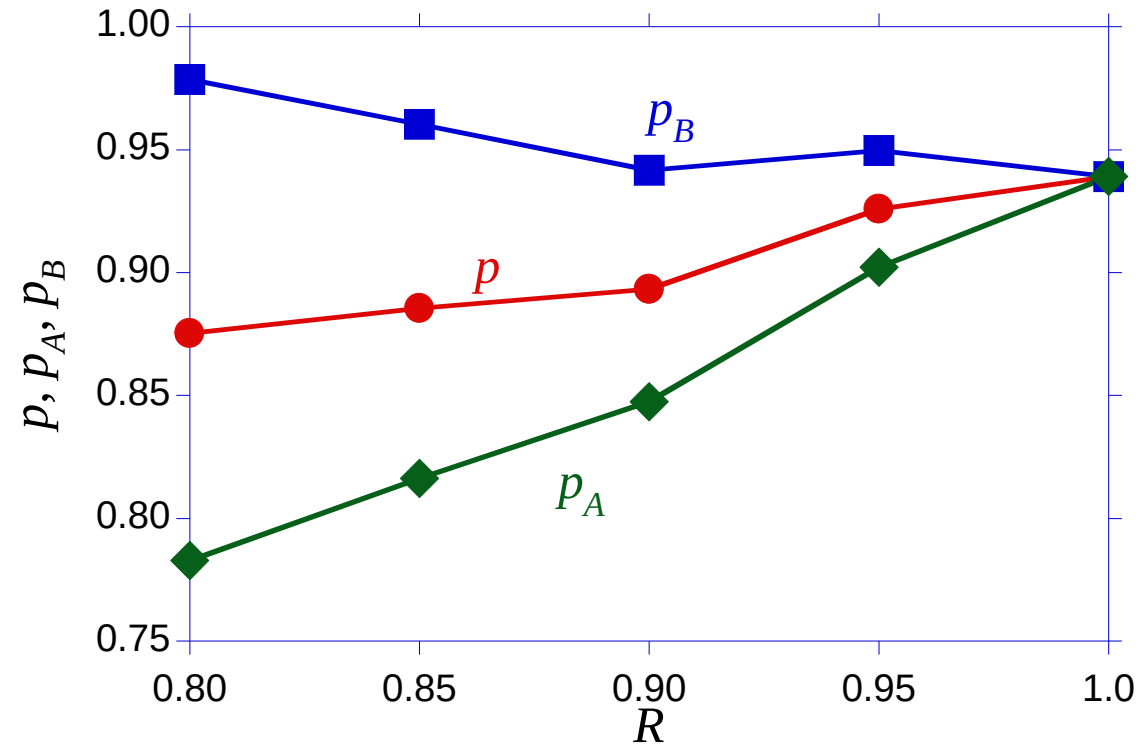

**Figure C-1.** Extent of reaction – overall as well as of species *A* and *B*, as function of *R*.

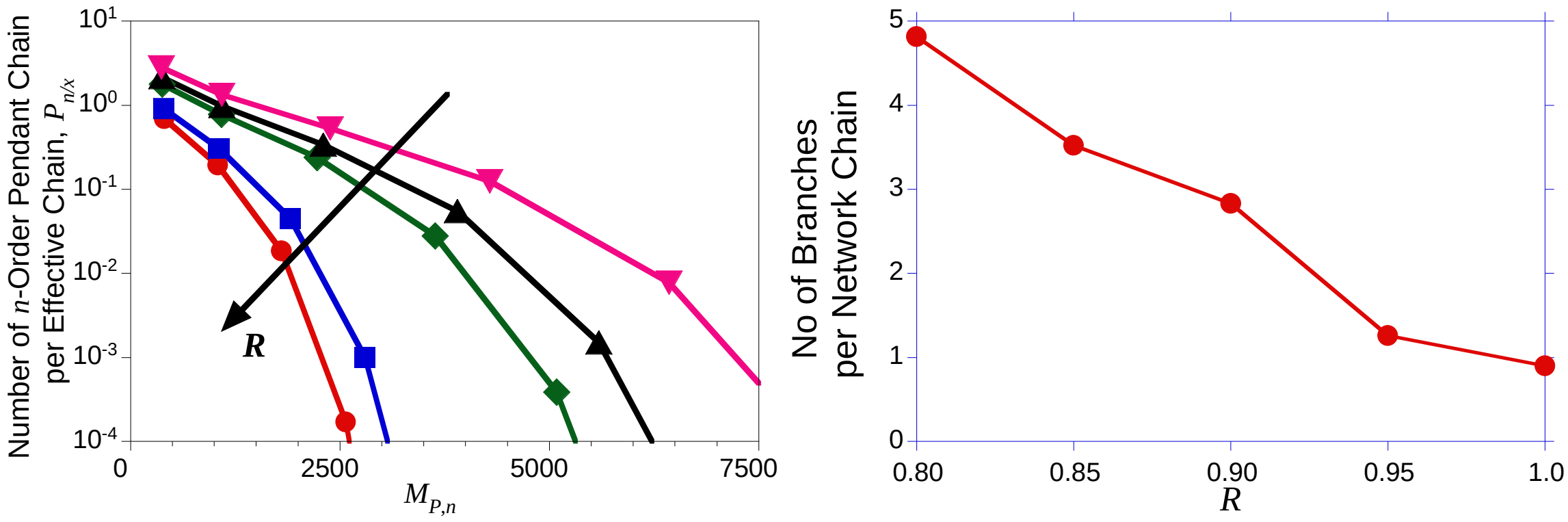

(a) (b)

**Figure C-2.** Pendant chains in the PBPU elastomer network. The lines only connect the data points, and are provided to aid the eye.(a) Distribution of Pendant Chain Lengths (b) No. of pendant chains per network chain

The sol fraction is then estimated via equations ( 25) to ( 27) for *R*=1. The difference between this value, and the corresponding sol fraction from swelling experiments (Figure B-2), can be considered to be the non-functional fraction of HTPB. The corresponding non-functional HTPB fractions for the other *R* values, can be estimated from the respective weight fractions of HTPB in the reaction mixture. On adding these to the Marsh model sol fractions (equations ( 27)), yields the sol fractions, which qualitatively follow those from swelling experiments (Figure C-3(a)). The pendant masses fractions via eqn. ( 43), are also plotted in Figure C-3(a).

Finally, we determine the bulk $G_X$, (Figure C-3(b)), by effectively applying eqn (15), recognizing that $M_X$ in this method corresponds to the affine model, $G_X = N\,\mathrm{k}T$. There is fair correspondence between the Marsh model results and the stress-elongation parameters, with excellent match for *R*=0.8, and the percolation limit being *R*~0.76.

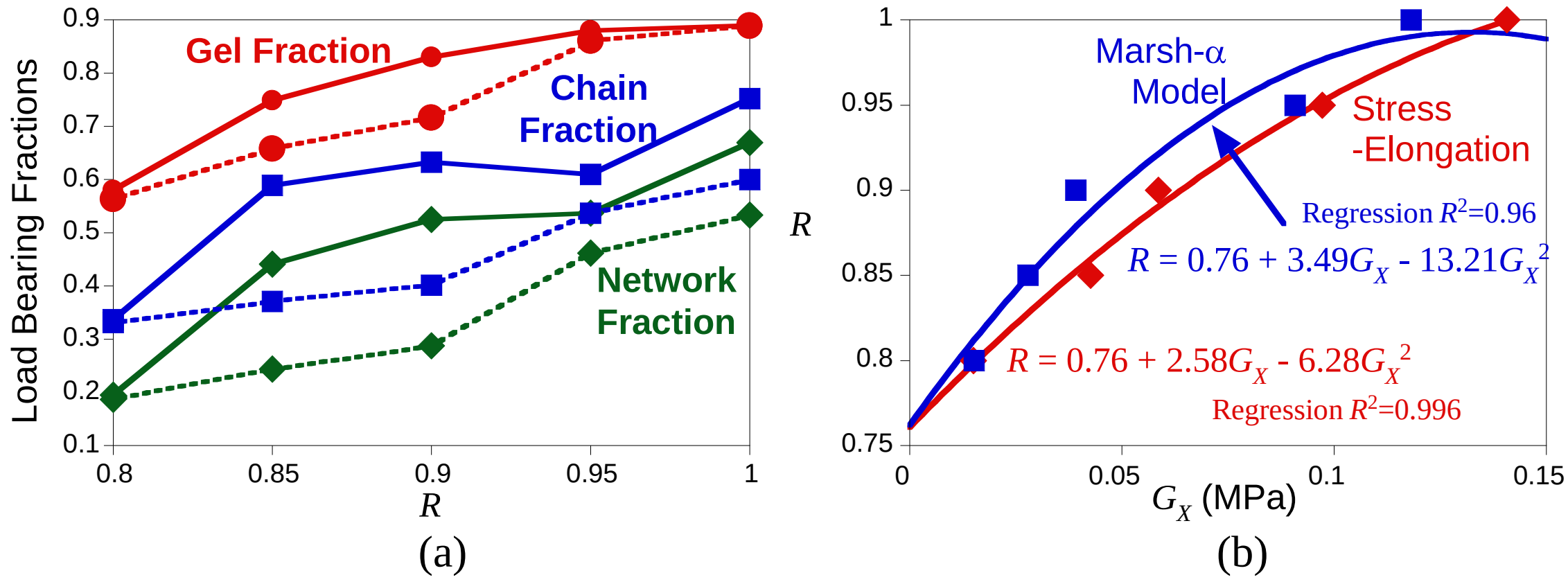

**Figure C-3.** Comparison between Marsh-α model and Stress Elongation Modeling (a) Load Bearing Fractions (Figure 8) (b) Crosslink Moduli to percolation (Figure 7).